\documentclass[10pt, 3p]{elsarticle}

\usepackage[english]{babel}
\usepackage[utf8]{inputenc}
\usepackage{graphics}
\usepackage{graphicx}
\usepackage{upgreek}
\usepackage{caption}
\usepackage{subcaption}
\usepackage{amsmath,amssymb,esint}
\usepackage{multirow}
\usepackage[usenames,x11names]{xcolor}
\usepackage[colorlinks]{hyperref}

\newcommand{\reviewerone}[1]{{\color{black}{#1}}}
\newcommand{\reviewertwo}[1]{{\color{black}{#1}}}
\newcommand{\reviewerthree}[1]{{\color{black}{#1}}}
\newcommand{\general}[1]{{\color{black}{#1}}}

\biboptions{sort&compress,authoryear}
\providecommand{\eprint}[2][]{arXiv:#2}

\journal{International Journal of Multiphase Flow}

\begin{document}

\begin{frontmatter}

\title{\reviewerthree{A conditional quadrature method of moments for populations of oblate spheroidal particles}}

\author[add1]{Mohamed Amine Bouguezzoul}
\author[add1]{Bianca Viggiano}
\author[add2]{Bruno Blais}
\author[add1]{Fabian Denner\corref{cor1}}

\ead{fabian.denner@polymtl.ca}

\address[add1]{Department of Mechanical Engineering, Polytechnique Montréal, Montréal, H3T 1J4, Québec, Canada}
\address[add2]{Department of Chemical Engineering, Polytechnique Montréal, Montréal, H3T 1J4, Québec, Canada}

\cortext[cor1]{Corresponding author: }

\begin{abstract}
Non-spherical particle populations are encountered in numerous industrial processes, where particle size, shape, and velocity jointly influence transport, breakage, and classification dynamics. The resulting coupled distributions of geometric and kinematic particle properties evolve within a high-dimensional phase space whose accurate representation remains a significant challenge. We present a conditional quadrature method of moments (CQMOM) for oblate spheroidal particles described by a five-dimensional phase space that includes their size, aspect ratio, and velocity. The dimensionality of the problem is reduced by assuming that the particles continuously adopt their drag-maximizing orientation. In the solution algorithm, the resulting conditional quadrature representation of the particle phase space is combined with a Strang splitting of the continuous and discontinuous \reviewerthree{mathematical operators} governing the moment transport. We \general{verify} this CQMOM framework against Monte Carlo simulations \general{of the same particle model} for representative test cases of increasing complexity, including drag, breakage, recoil, and attrition\general{, at different Stokes numbers. With two nodes per coordinate, the particle number, the mean size, and the mean aspect ratio agree with the Monte Carlo solution within its statistical uncertainty, and the velocity statistics to within 7\,\%. Time-step and quadrature resolution studies identify the conditional velocity quadrature as the main source of the remaining deviations, for which we discuss potential remedies. Overall, the results demonstrate that the conditional quadrature representation preserves the geometric-kinematic correlations of the particle population in a five-dimensional phase space, at a cost more than 60 times lower than that of a single Monte Carlo realization.}
\end{abstract}

\begin{keyword}
Quadrature-based moment methods \sep Conditional quadrature method of moments \sep Population balance modelling \sep Non-spherical particles\\~\\
\textcopyright~2026. This manuscript version is made available under the CC-BY 4.0 license. \url{http://creativecommons.org/licenses/by/4.0/}
\end{keyword}

\end{frontmatter}

\section{Introduction}

Disperse multiphase flows, involving a disperse phase composed of a large number of particles, droplets, or bubbles, are central to a wide variety of industrial and environmental processes and phenomena \citep{Balachandar2024}. Examples include the dynamics of liquid sprays \citep{Veron2015,Lefebvre2017}, cavitation \citep{Bryngelson2019, Coulombel2024}, \reviewertwo{mass transfer in stirred tank reactors \citep{Petitti2013}}, drug delivery \citep{vanWachem2017, Capecelatro2022}, crystallization processes \citep{Marchisio2026}, mineral processing \citep{Rajendran2023, Zhang2023h}, solid-liquid mixing \citep{Blais2016a}, and powder production \citep{Soong2023,Priyadarshi2024}. In most practical applications, the spatial and temporal scales of disperse multiphase systems span many orders of magnitude, while the disperse phase is typically polydisperse, meaning that its constituent elements exhibit a variability in properties such as size, shape, temperature, or velocity.

Two principal modelling paradigms are commonly employed to simulate disperse multiphase flows \citep{Balachandar2020a,Schneiderbauer2024}: Euler-Lagrange and Euler-Euler methods. In Euler-Lagrange methods \citep{Capecelatro2013}, individual particles, droplets, or bubbles are tracked explicitly together with their properties. Although very accurate, Euler-Lagrange methods become prohibitively expensive to deploy for industrial-scale systems due to the large number of disperse elements involved. In contrast, Euler-Euler methods \citep{VanWachem2023} treat the disperse phase as a continuum, resulting in computational costs that are largely independent of the number of disperse elements. However, classical Euler-Euler methods typically assume the disperse phase to be monodisperse and monokinetic, and therefore cannot adequately represent the broad distributions of particle properties encountered in real applications. Accurate and scalable simulations of polydisperse multiphase flows consequently require descriptions capable of capturing the essential characteristics of disperse populations without resolving every individual element. Population balance equations (PBEs) provide such a description by representing the properties of the particle population by a number density function (NDF) defined over an appropriate phase space \citep{Rigopoulos2024}.

The quadrature method of moments (QMOM) \citep{McGraw1997, Marchisio2013} offers an efficient framework for solving PBEs describing the evolution of a polydisperse population by its NDF. Rather than solving directly for the NDF, QMOM transports a finite set of low-order moments that provide a compact statistical description of the population. In the context of multiphase flows, these moments are directly related to physical quantities such as mass and momentum, whereby the corresponding moment transport equations represent the governing conservation laws of the disperse phase \citep{Fox2024, Heylmun2021, Patel2019a}, which in turn can be used as the basis for Euler-Euler methods \citep{Passalacqua2010, Li2022c, Kong2020, Li2019b}. However, the moment transport equations are, in general, not available in closed form, as they depend on higher-order or fractional-order moments. QMOM addresses this closure problem by approximating the NDF by a discrete quadrature representation composed of weighted Dirac delta functions. The quadrature weights and nodes are determined through a moment-inversion procedure rooted in the theory of orthogonal polynomials \citep{McGraw1997,Press1990}, whereby the transported low-order moments are used to reconstruct a Gaussian quadrature that \reviewerthree{reproduces them, provided that the moments are realizable}. Once reconstructed, the quadrature provides a compact representation of the distribution from which higher-order and fractional-order moments can be approximated and the moment transport equations can be closed. In this way, QMOM replaces the direct solution of a PBE by the transport of a relatively small number of moments, providing a computationally efficient alternative to sectional methods \citep{Li2020, Shiea2020} for the discrete representation of NDFs.

While QMOM works well for the solution of a single or uncoupled NDF, many disperse multiphase flows involve several coupled particle properties, including particle size, shape, velocity, composition, or temperature. Predicting such systems with a multidimensional phase space accurately requires not only the evolution of the marginal distributions of individual properties, but also the preservation of the coupling between them. \reviewertwo{Several quadrature-based approaches have been developed to this end \citep{Marchisio2013, Passalacqua2023}. The direct quadrature method of moments (DQMOM) \citep{Marchisio2005}, for instance, transports the weights and weighted nodes of a multivariate quadrature instead of the moments, which are then controlled only through the source terms of the weights and nodes. The conditional quadrature method of moments (CQMOM) \citep{Yuan2011}, in contrast, retains the moments, the conserved quantities of the disperse phase, as transported variables and} represents a multidimensional joint NDF as a hierarchy of conditional distributions, allowing coupled phase-space variables to be described within a moment-based framework. For instance, the joint NDF of particle size and velocity in a two-dimensional phase space can be expressed as the product of a marginal size distribution and a velocity distribution conditioned on size, thereby preserving the physical dependency of the particle velocity on the particle size. To this end, CQMOM constructs quadrature approximations for both the marginal and conditional distributions, resulting in a multidimensional quadrature representation of the underlying joint NDF. For particle-laden flows, this allows, for instance, to predict trajectory crossing of dilute or collision-less particle systems \citep{Yuan2011,Patel2019a}.

In recent years, several new QMOM-based methods have been proposed with the aim of improving the mathematical and numerical properties of quadrature-based moment methods. Notable examples include the hyperbolic formulations HyQMOM \citep{Fox2022,Morin2025} and CHyQMOM \citep{Fox2018,Patel2019a}, which are designed to ensure hyperbolicity of the moment transport equations, \reviewertwo{the extended quadrature method of moments (EQMOM) \citep{Yuan2012}, which reconstructs a continuous NDF as a sum of kernel density functions,} and the generalized quadrature method of moments (GQMOM) \citep{Fox2023a}, which exploits prior knowledge of the underlying distribution family (e.g., a log-normal distribution) to improve moment reconstruction. \reviewerthree{If the underlying distribution belongs to the assumed distribution family, GQMOM also reproduces moments of higher order than the transported low-order moments, improving accuracy and robustness.} Significant advances have also been made in the numerical integration of the moment transport equations. Following the derivation of the moment system, the evolution of the transported moments is commonly decomposed into advection in physical space, continuous evolution in phase space, and discontinuous evolution in phase space \citep{Yuan2011, Heylmun2021}. While the advection in physical space can generally be treated using conventional computational fluid dynamics techniques for advection equations \citep{Vikas2011, Laurent2017, Shiea2020a}, the evolution of the population and its properties in phase space lies at the core of QMOM and CQMOM methodologies. \reviewerthree{Continuous processes, such as drag or evaporation, enter the PBE through advection in phase space, whereas discontinuous processes, such as collisions and breakage, enter through integral source terms that relocate number density between distant states in phase space. The fundamentally different mathematical nature of these operators has} motivated the widespread adoption of operator-splitting strategies \citep{Yuan2011, Yuan2012, Vikas2011,Nguyen2016, Li2022c, Heylmun2021, Posey2026,Li2019a,Patel2019a}, which allow numerical methods tailored to each process to be employed and improve the accuracy, robustness, and efficiency of the overall solution.

\begin{figure}
    \centering
    \includegraphics[width=0.7\linewidth]{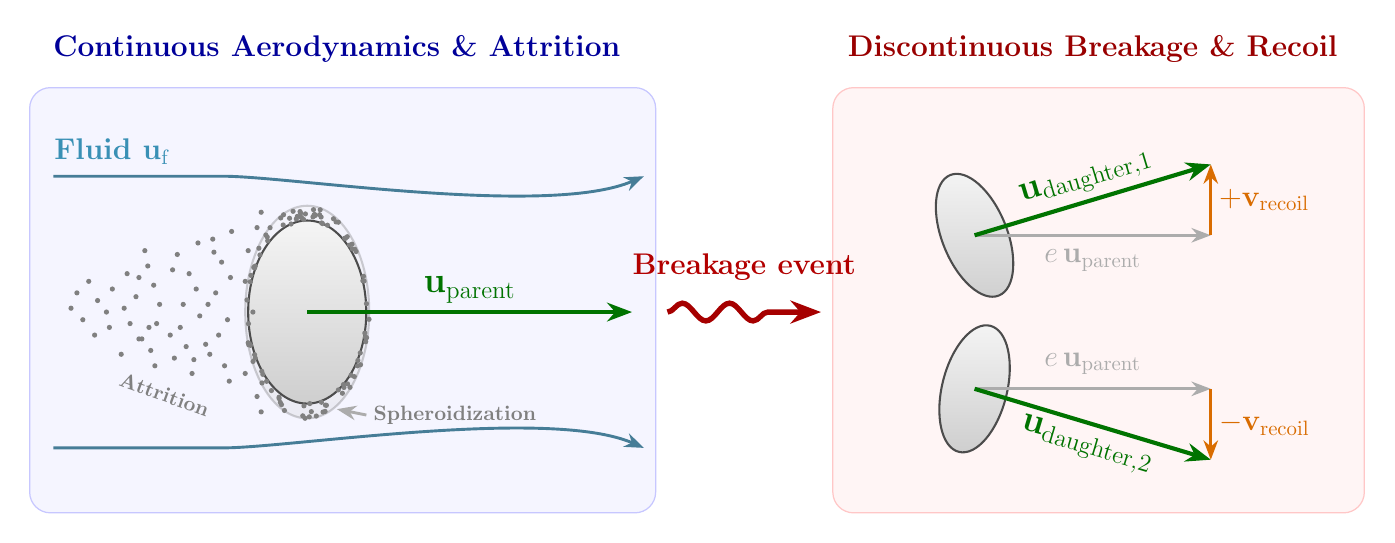}
    \caption{Schematic illustration of the coupled geometric and kinematic particle dynamics targeted by the proposed CQMOM framework for non-spherical particles. Aerodynamic drag and attrition induce a continuous evolution of particle size, shape, and velocity, while breakage events generate discontinuous changes in particle geometry and velocity.}
    \label{fig:roadmap}
\end{figure}

Despite its success in applications involving spherical particles, a CQMOM framework for non-spherical particle populations has not yet been proposed in the literature. Non-spherical particles are ubiquitous in particle technology, where particle size and shape strongly influence transport, collision, breakage, and classification dynamics. Accurately describing such systems requires the simultaneous representation of coupled size, shape, and velocity distributions, leading to a high-dimensional phase space and significantly increasing the complexity of both the underlying PBE and its numerical solution. To address this challenge, the present work proposes a five-dimensional CQMOM framework for non-spherical particles represented by oblate spheroids and characterized by their size, aspect ratio, and velocity.  As illustrated in Figure~\ref{fig:roadmap}, the targeted particle populations are subject to the simultaneous action of continuous aerodynamic and attrition-driven processes, which induce smooth evolution of the particle geometry and velocity, as well as breakage events that generate discontinuous changes in particle size, shape, and velocity. The objective of the proposed CQMOM framework, which we \reviewerthree{verify against Monte Carlo simulations of representative test cases}, is to accurately capture the coupled evolution of these geometric and kinematic particle properties within a computationally tractable moment-based description.

The remainder of this manuscript is organized as follows. Section \ref{sec:governing} introduces the equations governing the considered non-spherical particle populations, and Section \ref{sec:model} presents the employed model for \reviewerthree{oblate spheroidal} particles and discusses its underpinning assumptions. The CQMOM developed to model the evolution of the five-dimensional phase space is introduced in Section \ref{sec:cqmom} and the solution algorithm is presented in Section \ref{sec:solution}. The complete CQMOM framework for populations of \reviewerthree{oblate spheroidal} particles is tested \general{in three test cases of increasing complexity in Section~\ref{sec:results}, and Section~\ref{sec:discussion} discusses the dependence of the results on the numerical parameters, the computational cost, and the scope of the method}. The conclusions of this work are presented in Section \ref{sec:conclusions}.

\section{Governing equations}
\label{sec:governing}

The properties of a polydisperse particle population can be described by an NDF $n(t,\mathbf{x},\boldsymbol{\xi})$, where $t$ is time,  $\mathbf{x} \equiv (x,y,z)$ denotes the physical space, and $\boldsymbol{\xi}$ is the phase-space vector that contains all relevant particle properties, such as size and velocity. The evolution of this NDF is governed by a PBE \citep{Rigopoulos2024},
\begin{equation}
    \frac{\partial n}{\partial t} + \nabla_{\mathbf{x}} \cdot (n \mathbf{u}) + \nabla_{\boldsymbol{\xi}} \cdot (n \dot{\boldsymbol{\xi}} ) = {S}, \label{eq:pbe}
\end{equation}
where $\mathbf{u}$ denotes the physical-space velocity of the particles, $\dot{\boldsymbol{\xi}}$ is the phase-space velocity associated with continuous changes of the particle properties, and $S$ represents discontinuous source terms. 

The direct numerical solution of Eq.~(\ref{eq:pbe}) rapidly becomes prohibitively expensive as the dimensionality of the phase space increases \citep{Shiea2020}. Moment methods address this challenge by replacing the evolution of the NDF with the evolution of a finite set of its low-order moments. The moment of order $\mathbf{k}$ of the NDF $n(t,\mathbf{x},\boldsymbol{\xi})$ is defined as
\begin{equation}
M_{\mathbf{k}}(t,\mathbf{x}) = \int_{\Omega} \boldsymbol{\xi}^{\mathbf{k}} \, n(t,\mathbf{x},\boldsymbol{\xi}) \,\text{d}\boldsymbol{\xi}, \label{eq:moment_def}
\end{equation}
where $\mathbf{k}=(k_1,\ldots,k_{N_\xi})$, $\Omega$ denotes the phase space, and $N_\xi$ is the number of phase-space variables. Multiplying Eq.~\eqref{eq:pbe} by $\boldsymbol{\xi}^{\mathbf{k}}$ and integrating over the phase space $\Omega$ yields the moment transport equation
\begin{equation}
\frac{\partial M_{\mathbf{k}}}{\partial t} + \nabla_{\mathbf{x}} \cdot \left( \int_{\Omega} \boldsymbol{\xi}^{\mathbf{k}} \, n \, \mathbf{u} \, \text{d}\boldsymbol{\xi} \right) = C_{\mathbf{k}} + D_{\mathbf{k}},
\label{eq:mte_full}
\end{equation}
where the second term on the left-hand side represents advection in physical space, the continuous processes are described by
\begin{equation}
C_{\mathbf{k}} = -\int_{\Omega} \boldsymbol{\xi}^{\mathbf{k}} \, \nabla_{\boldsymbol{\xi}} \cdot \left( n\dot{\boldsymbol{\xi}} \right)  \,\text{d}\boldsymbol{\xi}, \label{eq:continuous_processes}
\end{equation}
and the discontinuous processes are described by
\begin{equation}
D_{\mathbf{k}}
= \int_{\Omega} \boldsymbol{\xi}^{\mathbf{k}} S \,\text{d}\boldsymbol{\xi}.
\label{eq:discontinuous_processes}
\end{equation}
The evaluation of these terms generally requires information about the NDF beyond the transported moments, giving rise to the moment-closure problem addressed by quadrature-based moment methods, such as QMOM and CQMOM.

The moment transport equation, Eq.~(\ref{eq:mte_full}), naturally decomposes into three operators corresponding to the advection in physical space, the continuous evolution in phase space, and the discontinuous evolution in phase space. The advection in physical space (i.e., through the computational domain) is generally treated using conventional computational fluid dynamics techniques, such as finite-volume methods \citep{Vikas2011}. Since the objective of the present work is to develop a numerical method for the phase-space dynamics of non-spherical particle populations, the advection in physical space is neglected and only the evolution in phase space is considered, such that $M_\mathbf{k}(t)$ is governed by 
\begin{equation}
\frac{\text{d} M_{\mathbf{k}}}{\text{d} t} = C_{\mathbf{k}} + D_{\mathbf{k}}
\label{eq:mte}
\end{equation}
through a combination of aerodynamic forces, surface attrition, recoil, and breakage. \reviewerone{The particle populations are assumed to be dilute and collisionless, so that interactions between particles, including collisions, are not considered.} Collectively, these mechanisms induce a coupled evolution of particle size, shape, and velocity through both continuous and discontinuous phase-space dynamics, providing a demanding test of the ability of CQMOM to preserve correlations between the particle properties and accurately represent their evolution. The particle-scale models used to define the phase-space velocity $\dot{\boldsymbol{\xi}}$ and the source term $S$, as well as the definition of the considered phase space $\boldsymbol{\xi}$, are presented in Section \ref{sec:model}.

Throughout this work, the terms \textit{continuous process} and \textit{discontinuous process} refer to \reviewerthree{the mathematical nature of the operators by which the processes are represented in Eq.~\eqref{eq:pbe}. Continuous processes, such as aerodynamic drag and attrition, enter the phase-space advection term $\nabla_{\boldsymbol{\xi}} \cdot (n \dot{\boldsymbol{\xi}})$, a local differential operator that transports the NDF continuously along characteristics, which are the particle trajectories in phase space. Discontinuous processes, such as breakage, enter the source term $S$ as birth-death terms, nonlocal integral operators that relocate number density between distant states in phase space. This classification refers to the representation adopted in the PBE at the considered modelling scale. Attrition, for instance, is represented as a continuous process after coarse-graining. The distinction determines the numerical treatment of the two operators, as discussed in Section~\ref{sec:solution}.}

\section{Model for non-spherical particles}
\label{sec:model}

This section presents the particle-scale model used to describe the evolution of non-spherical particles within the proposed CQMOM framework. \reviewerthree{The particle model describes dilute populations of rigid oblate spheroids with a quasi-steady orientation. It serves to define a demanding but well-controlled test problem for the CQMOM framework, in which continuous and discontinuous processes act simultaneously on the geometric and kinematic coordinates of the phase space. However, the computational framework itself does not depend on the particular closures: it only requires the phase-space velocity, the breakage rate, and the daughter distribution to be evaluated at the quadrature nodes, so that the model for the aerodynamic forces as well as the deliberately simple models of attrition, breakage, and recoil employed below can each be replaced by any other closure, for instance one suited to a different particle class or Reynolds-number range. Since the Monte Carlo reference solutions in Section~\ref{sec:results} employ the same closures, the comparison verifies the numerical method for this particle model, but does not validate the particle model itself.}

One of the principal strengths of quadrature-based moment methods, such as QMOM and CQMOM, is their ability to directly incorporate particle-scale physics into the description of the disperse phase \citep{Marchisio2013}. While the transported moments characterize the statistical evolution of the particle population, the reconstructed quadrature provides access to representative particle states at which particle-scale models can be evaluated. Consequently, models derived from experiments, theory, or fully resolved simulations can be incorporated naturally into the moment equations.

\begin{figure}
    \centering
    \includegraphics[scale=0.9]{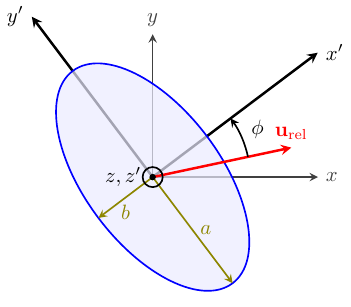}
    \caption{Geometry of an oblate spheroidal particle. The particle is characterized by its equatorial semi-major axis $a$ and polar semi-minor axis $b$. The particle-fixed reference frame ($x',y',z'$) may rotate relative to the global frame ($x,y,z$). The angle of incidence $\phi$ is defined as the angle between the \reviewerthree{relative velocity $\mathbf{u}_\text{rel}$ between fluid and particle} and the particle symmetry axis $x'$.}
    \label{fig:spheroid}
\end{figure}

The non-spherical particles considered in this study are assumed to be oblate spheroids, each characterized by their equatorial semi-major axis $a$ and their polar semi-minor axis $b \leq a$, shown in Figure \ref{fig:spheroid}. The particle size and shape are described by the volume-equivalent diameter $d = \sqrt[3]{8a^2b}$ and the aspect ratio $0 < \chi = b/a \leq 1$ of the oblate spheroid, respectively. An aspect ratio of $\chi =1$ represents a sphere, whereas $\chi \rightarrow 0$ describes a disk. Following \citet{Ouchene2020}, the reference area for aerodynamic forces is $A_\text{ref} = \pi d^2/4$ and the angle of incidence $\phi$ is defined as the angle between the \reviewerthree{relative velocity $\mathbf{u}_\text{rel}$, defined below,} and the polar minor axis (i.e., the axis of symmetry) of the oblate spheroid, as illustrated in Figure \ref{fig:spheroid}.

\subsection{Equations of motion}
\label{sec:eom}

The equations of motion of an arbitrarily shaped solid particle follow from Newton’s second law and the conservation of angular momentum as
\begin{align}
    m \frac{\text{d} \mathbf{u}}{\text{d}t} = \mathbf{F} ,\label{eq:translation} \\
    \mathbf{J} \cdot \frac{\text{d}\boldsymbol{\omega}}{\text{d}t} + \boldsymbol{\omega} \times (\boldsymbol{\omega} \cdot \mathbf{J})  = \mathbf{T}, \label{eq:rotation}
\end{align}
where $\mathbf{u} \equiv (u , v , w)$ is the translational velocity of the particle, $\boldsymbol{\omega} \equiv (\omega_{x'} , \omega_{y'} , \omega_{z'})$ is the angular velocity of the particle, $m$ is the particle mass, $\mathbf{J}$ is the inertia tensor of the particle, while $\mathbf{F}$ and $\mathbf{T}$ are the forces and torques acting on the particle. Note that all quantities in Eq.~\eqref{eq:rotation} are defined in the particle frame of reference ($x', y', z'$).

For axisymmetric particles, such as the considered oblate spheroids, the drag and shape-induced lift forces acting on each particle are generally defined as \citep{Ouchene2020}
\begin{align}
    \mathbf{F}_\text{D} &= \frac{\rho_\text{f}}{2} \, A_\text{ref} \, C_\text{D}(\phi, \chi, \text{Re}) \, \| \mathbf{u}_\text{rel} \| \, \mathbf{u}_\text{rel} \label{eq:FD},\\
    \mathbf{F}_\text{L} &=  \frac{\rho_\text{f}}{2} \, A_\text{ref} \, C_\text{L}(\phi, \chi, \text{Re}) \, \| \mathbf{u}_\text{rel} \|^2 \, \mathbf{e}_\text{L},
\end{align}
respectively, where $\rho_\text{f}$ is the fluid density, $\mathbf{u}_\text{rel}=\mathbf{u}_\text{f} - \mathbf{u}$ is the relative velocity between the fluid and the particle, and $\mathbf{e}_\text{L}$ is the unit vector normal to $\mathbf{u}_\text{rel}$ in the plane spanned by $\mathbf{u}_\text{rel}$ and the symmetry axis of the particle. 
The pitching torque acting on a non-spherical particle can be formulated in a similar manner as \citep{Ouchene2020, Zastawny2012c}
\begin{align}
    \textbf{T} = \frac{\rho_\text{f}}{2} \, A_\text{ref} \, d \, C_\text{T}(\phi, \chi, \text{Re}) \, \| \mathbf{u}_\text{rel} \|^2 \, \mathbf{e}_\text{T} ,  \label{eq:TP}
\end{align}
where $\mathbf{e}_\text{T}$ is the unit vector normal to this plane\reviewerthree{, oriented such that the torque turns the symmetry axis towards $\mathbf{u}_\text{rel}$}, while the rotational torque follows from the relative angular velocity between particle and fluid. 
The drag, lift, and torque coefficients ($C_\text{D}$, $C_\text{L}$, $C_\text{T}$) of a particle with spheroidal shape are a function of the angle of incidence $\phi$ of the flow on the particle, the particle aspect ratio $\chi$, and the particle Reynolds number
\begin{equation}
    \text{Re} = \frac{\rho_\text{f} \, d \, \| \mathbf{u}_\text{rel} \|}{\mu_\text{f}},
\end{equation}
where $\mu_\text{f}$ is the dynamic viscosity of the fluid. 

According to Eqs.~\eqref{eq:FD}-\eqref{eq:TP}, the aerodynamic particle state can be described by a seven-dimensional phase space, $\boldsymbol{\xi} = (d, \chi, u, v, w, \phi, \omega_{z'})$. However, this formulation does not consider that the particle may also pitch out-of-plane with respect to the equatorial major axis, a common assumption in studies on the aerodynamic forces of non-spherical axisymmetric particles \citep{Ouchene2020, Zastawny2012c, Sanjeevi2018, Cheron2024, Gorges2025}, which would add the angular velocity $\omega_{y'}$ and the corresponding orientation angle to the phase space. Additionally, accounting for the rotation around the polar minor axis of the oblate spheroid by including $\omega_{x'}$ \reviewerthree{and the corresponding roll angle, which completes the description of the orientation of the particle frame $(x',y',z')$,} is required if rotational torque is fully accounted for \citep{Rosendahl2000}.
\reviewerthree{Resolving the orientation and rotational dynamics of the particles would therefore extend the phase space to as many as eleven dimensions. Applying CQMOM to such an $11$-dimensional phase space would require tracking $6142$ moments for a quadrature with $2$ nodes per coordinate or $442866$ moments for a quadrature with $3$ nodes per coordinate. This would be neither computationally tractable nor numerically robust, since every additional level of the conditional hierarchy multiplies the number of branches and reduces their weights, and the reconstruction of such light branches is the most fragile part of the method, as further discussed in Section~\ref{sec:discussion}.} 

To avoid explicitly resolving the rotational dynamics, we neglect rotational torque and assume that, as a result of the acting pitching torque, the non-spherical particles maximize their drag by aligning their polar minor axis with the \reviewerthree{relative velocity $\mathbf{u}_\text{rel}$}, resulting in a quasi-steady orientation (QSO) of each particle with $\phi=0^\circ$. Drag maximization of non-spherical objects is common in nature \citep{Hoerner1965}, and non-spherical particles in turbulent flows have been observed to tend towards a drag-maximizing orientation \citep{VanWachem2015}. \reviewerthree{Nevertheless, this quasi-steady, drag-maximizing orientation of the particles is an explicit assumption made to simplify the particle model. The price of the reduced phase space is that the particle model does not represent unsteady orientation regimes, such as the fluttering and tumbling of freely falling disks and oblate bodies, which occur at Reynolds numbers of order 100 and above \citep{Ern2012, Zhou2017, Tinklenberg2023}, nor orientation-dependent forces in non-uniform flows, such as shape-induced lift in velocity gradients.} As a consequence of the QSO assumption, the state of the oblate spheroidal particles is described by the five-dimensional phase space $\boldsymbol{\xi} = (d, \chi, u, v, w)$, containing particle size $d$, aspect ratio $\chi$, and velocity $\mathbf{u} = (u, v, w)$.

\subsection{Aerodynamic forces}
\label{sec:forces}

Analytical and empirical models for the drag coefficient of axisymmetric non-spherical particles are typically based on the formulation
\begin{align}
    C_\text{D} &= C_{\text{D},\phi=0^\circ} + (C_{\text{D},\phi=90^\circ} - C_{\text{D},\phi=0^\circ}) \sin^\beta \phi,
\end{align}
with $\beta =2$ \citep{Happel1983,  Sanjeevi2018, Ouchene2020} or $\beta =3$ \citep{Mando2010, Zastawny2012c}. Similarly, models for the shape-induced lift coefficient of such particles are usually based on \citep{Mando2010, Zastawny2012c}
\begin{align}
    C_\text{L} = C_{\text{D}} \, \sin^2 \phi \, \cos \phi
\end{align}
or \citep{Ouchene2020, Sanjeevi2018}
\begin{align}
    C_\text{L} &= (C_{\text{D},\phi=90^\circ} - C_{\text{D},\phi=0^\circ}) \sin \phi \, \cos \phi.
\end{align}
Under the QSO assumption with $\phi=0^\circ$, these models simplify to $C_\text{D} = C_{\text{D},\phi=0^\circ}$ and $C_\text{L} = 0$. Hence, only the drag coefficient associated with an alignment of the polar minor axis and the \reviewerthree{relative velocity} vector ought to be considered, representing the drag-maximizing orientation of an oblate spheroid, and shape-induced lift can be neglected.

Neglecting buoyancy and unsteady forces acting on the particles, which are common assumptions for solid particles in a gas \citep{VanWachem2023,Fox2024}, and applying the QSO assumption, the acceleration of the considered oblate spheroids is defined by
\begin{equation}
    \frac{\text{d}\mathbf{u}}{\text{d}t} = \frac{\mathbf{F}_\text{D}}{m} + \mathbf{g}, \label{eq:particle_acceleration}
\end{equation}
where the drag force $\mathbf{F}_\text{D}$ is given by Eq.~\eqref{eq:FD} and $\mathbf{g}$ denotes the gravitational acceleration. In this work we use the drag correlation for oblate spheroids proposed by \citet{Ouchene2020}, for which the drag coefficient is defined as 
\begin{equation}
    C_\text{D}(\phi=0^\circ,\chi,\text{Re}) = \frac{24}{\text{Re}} \left(K_{\phi=0^\circ} + 0.15 \, \chi^{95.91} \, \text{Re}^{0.687} + 0.2927 \, (1-\chi)^{0.4374} \, \text{Re}^{0.7512}\right),
\end{equation}
where
\begin{equation}
    K_{\phi=0^\circ} = \frac{8}{3 \, {\chi^{1/3}}} \left[ \frac{2\chi}{1-\chi^2} + \frac{2 \, (1-2\chi^2)}{\left(1-\chi^2\right)^{3/2}} \, \tan^{-1} \left(\frac{\sqrt{1-\chi^2}}{\chi}\right) \right]^{-1}.
\end{equation}
\reviewerthree{This correlation was obtained for $0.2 \leq \chi \leq 1$ and $\text{Re} \leq 100$ \citep{Ouchene2020}; beyond this range, as for part of the particles considered in this study, it is extrapolated.} Note that the proposed framework is independent of the particular drag-force model employed and alternative correlations may be substituted to better represent specific Reynolds-number regimes.

\subsection{Attrition}
\label{sec:attrition}

Particle attrition is modelled as a continuous evolution in phase space acting on both the size and the aspect ratio of the particle. \reviewerthree{The attrition model is a phenomenological closure, chosen to exercise the coupling of continuous and discontinuous processes in phase space, rather than a predictive model of attrition.} Following empirical observations that attrition rates scale with the kinetic energy of the particle motion relative to the fluid \citep{Reppenhagen2000, Ghadiri2002}, the particle diameter evolves according to
\begin{equation}
\reviewerthree{\frac{\text{d}d}{\text{d}t} = -k_{\text{attrition}} \, \frac{\left\| \mathbf{u}_{\text{rel}} \right\|^2}{1 + \frac{\left\| \mathbf{u}_{\text{rel}} \right\|^2}{U_s^2}} \, \max\!\left[0, \tanh\!\left(\frac{d - d_\text{min}}{d_s}\right)\right],} \label{eq:d_continuous}
\end{equation}
where $k_{\text{attrition}}$ is the attrition coefficient\reviewerthree{, in s/m, and $\left\| \mathbf{u}_{\text{rel}} \right\|$ is the Euclidean norm of the relative velocity. The rate saturates at relative speeds above $U_s$ and vanishes smoothly as the diameter approaches a minimum diameter $d_\text{min}$, over the width $d_s$, so that the diameter remains positive and cannot increase. 
Eq.~\eqref{eq:d_continuous} represents attrition in a simplified form as an erosion driven by the slip velocity, whereby the lumped coefficient $k_\text{attrition}$ includes the dependencies on the material properties of the particle, on collisions and contact conditions, and on the properties of the fluid.} Simultaneously, the particle shape relaxes gradually toward a spherical shape,
\begin{equation}
\frac{\text{d}\chi}{\text{d}t} = \frac{1-\chi}{\tau_{\text{shape}}},
\label{eq:chi_continuous}
\end{equation}
where $\tau_{\text{shape}}$ is the characteristic timescale of this shape relaxation.

\subsection{Breakage}
\label{sec:breakage}

To incorporate particle breakage into the moment framework, we consider a symmetric binary breakage, whereby a particle breaks into two equally sized daughter particles. The corresponding source term is given as \citep{Shiea2020, Marchisio2013}
\begin{equation}
S_{\text{break}}(\boldsymbol{\xi}) = \int_{\Omega}
\beta(\boldsymbol{\xi}|\boldsymbol{\xi}') \,  R_{\text{break}}  (\boldsymbol{\xi}') \, n(\boldsymbol{\xi}') \, \text{d}\boldsymbol{\xi}' - R_{\text{break}}(\boldsymbol{\xi}) \, n(\boldsymbol{\xi}), \label{eq:breakage_source}
\end{equation}
where $\beta(\boldsymbol{\xi}|\boldsymbol{\xi}')$ is the daughter
distribution function\general{, the number density of daughter particles formed in the state $\boldsymbol{\xi}$ by the breakage of a parent particle in the state $\boldsymbol{\xi}'$, which integrates to the number of daughter particles}.
The breakage frequency $R_{\text{break}}$ is prescribed by the size-dependent kernel \citep{Rigopoulos2024}
\begin{equation}
R_{\text{break}}(d)= k_{\text{break}} \,d^\gamma, \label{eq:Rbreak}
\end{equation}
with $\gamma=2$, ensuring that the breakage rate is proportional to the surface area of the particle. The coefficient $k_{\text{break}}$ is chosen based on the considered particle size such that multiple daughter generations are produced within the simulation time frame considered for the test cases presented in Section \ref{sec:results}. 
Given that each parent particle generates two daughter particles \general{of equal size and shape}, the daughter distribution function $\beta(\boldsymbol{\xi}|\boldsymbol{\xi}')$ is defined as
\begin{equation}
\general{\beta(\boldsymbol{\xi}|\boldsymbol{\xi}') = 2\,\delta \left(d - d_\text{daughter}(d')\right) \, \delta \left(\chi - \chi_\text{daughter}(\chi')\right) \, \mathcal{N}\left(\mathbf{u}; e \, \mathbf{u}', \sigma^2_\text{recoil} \mathbf{I}\right),}
\end{equation}
\general{where $d_\text{daughter}$ and $\chi_\text{daughter}$ as well as the parameters $e$ and $\sigma_\text{recoil}$ of the daughter velocity are defined below, and $\mathcal{N}(\mathbf{u}; e \, \mathbf{u}', \sigma^2_\text{recoil} \mathbf{I})$ denotes the probability density of the daughter velocity $\mathbf{u}$, a normal distribution with mean $e \, \mathbf{u}'$ and covariance matrix $\sigma^2_\text{recoil} \mathbf{I}$, that is, with each velocity component independently normally distributed around $e$ times the corresponding parent velocity component with variance $\sigma^2_\text{recoil}$. Without recoil, this density reduces to $\delta(\mathbf{u} - e \, \mathbf{u}')$.}
Conservation of the particle volume then yields
\begin{equation}
\general{d_{\text{daughter}}(d') = \frac{d'}{\sqrt[3]{2}}}, \label{eq:d_daughter}
\end{equation}
while the velocities of the two daughter particles are defined as
\begin{align}
\mathbf{u}_{\text{daughter}} &= e \, \general{\mathbf{u}'} \pm \mathbf{v}_\text{recoil}, \label{eq:v_recoil}
\end{align}
where $e$ is the fraction of the parent momentum inherited by the daughter particles and $\mathbf{v}_\text{recoil}$ is a superimposed recoil velocity. The recoil velocity is either zero ($\mathbf{v}_\text{recoil} = \mathbf{0}$) or sampled from a normal distribution, $\mathbf{v}_\text{recoil} \sim \mathcal{N}(\mathbf{0}, \sigma^2_\text{recoil} \mathbf{I})$. \general{Since the two daughter particles receive the recoil velocity with opposite signs, a breakage event with $e = 1$, as in the test cases of Section~\ref{sec:results}, conserves the momentum of the parent particle, whereas the recoil adds kinetic energy.}

To account for changes of the particle shape as a result of breakage, the aspect ratio of the daughter particles is prescribed either by
\begin{equation}
\general{\chi_{\text{daughter}}(\chi') = \chi' + \alpha \left(1-\chi'\right)}, \label{eq:chi_daughter_spherical}
\end{equation}
driving the particles toward more spherical shapes, or by
\begin{equation}
\general{\chi_{\text{daughter}}(\chi') = \alpha \, \chi'}, \label{eq:chi_daughter_disk}
\end{equation}
driving the particles toward increasingly oblate shapes, where $\alpha$ is a \general{shape parameter}.

\section{Conditional quadrature method of moments}
\label{sec:cqmom}

The direct numerical solution of the PBE introduced in Section~\ref{sec:governing} for the five-dimensional phase space identified for non-spherical particles in Section \ref{sec:eom} is prohibitively expensive due to the high dimensionality of the phase space. Instead, the present work employs the conditional quadrature method of moments (CQMOM) \citep{Yuan2011} to represent the particle population through a finite set of transported moments. In contrast to previous CQMOM formulations for spherical particles \citep{Heylmun2021,Li2022c}, the present work considers a phase space $\boldsymbol{\xi}$ containing particle size $d$, aspect ratio $\chi$, and velocity $\mathbf{u} = (u, v, w)$. 

\subsection{Phase-space representation}

The state of each particle is described by the phase-space coordinate $\boldsymbol{\xi} = (d, \chi, u, v, w)$ and the corresponding number density function for this five-dimensional phase space is $n = n(d, \chi, u, v, w)$. The moments of this number density function, following Eq.~\eqref{eq:moment_def}, are
\begin{equation}
    M_{ijklm} = \int_\Omega d^i \, \chi^j \,  u^k \,  v^l \,  w^m \, n \ \text{d}d \, \text{d}\chi \, \text{d}u \, \text{d}v \, \text{d}w, \label{eq:5d-moment}
\end{equation}
where $i$, $j$, $k$, $l$, and $m$ are non-negative integers and $\Omega$ denotes the phase space. The zeroth-order moment $M_{00000}$ corresponds to the number of particles, while higher-order moments characterize the size, shape, and velocity distributions as well as their mutual correlations. Rather than transporting the full number density function, CQMOM reconstructs an approximate representation of $n$ from a finite set of moments through a hierarchy of conditional quadratures, which permits the efficient representation of the joint size-shape-velocity distribution.

The ordering of the conditional hierarchy is not unique and influences both the numerical conditioning of the inversion procedure and the ability of the quadrature to represent correlations between phase-space variables. In general, a favourable strategy may be to order the phase-space variables according to their variance \citep{Fox2023a}, such that the variable with the largest variance forms the basis of the conditional hierarchy. In the present work, the variables are ordered according to \general{$d \rightarrow \chi \rightarrow w \rightarrow u \rightarrow v$}, reflecting the underlying particle dynamics. The particle aspect ratio is strongly coupled to the particle size through the breakage model, while the particle velocity depends on both the particle size and aspect ratio as a result of aerodynamic forces. Hence, the velocity distributions are conditioned on the particle properties that govern their evolution. \general{Among the velocity components, the horizontal components $u$ and $v$ are conditioned last, because their conditional distributions become nearly monokinetic as the particles follow the carrier flow. We revisit the conditioning order in Section~\ref{sec:discussion_resolution}.} 

\subsection{Conditional quadrature representation}

The principal idea of CQMOM is to represent a multidimensional number density function through a hierarchy of one-dimensional conditional distributions. For the five-dimensional phase space considered in the present work, the NDF is decomposed as
\begin{equation}
n(d,\chi,u,v,w) = n_{d}(d) \ n_{\chi|d}(\chi|d) \ \general{n_{w|d,\chi}(w|d,\chi) \ n_{u|d,\chi,w}(u|d,\chi,w) \ n_{v|d,\chi,w,u}(v|d,\chi,w,u)},
\label{eq:conditional_ndf}
\end{equation}
where $n_d$ denotes the marginal size distribution and the remaining terms represent successive conditional distributions. This decomposition allows statistical correlations between particle size, shape, and velocity to be represented, while reducing the multidimensional inversion problem to a sequence of one-dimensional inversions.

Each marginal or conditional distribution is approximated by a discrete quadrature representation \citep{Yuan2011}. The marginal size distribution is represented as
\begin{equation}
n_d(d) \approx \sum_{\alpha=1}^{N_d} \rho_\alpha \, \delta \!\left(d - d_\alpha \right),
\label{eq:size_quadrature}
\end{equation}
where \general{$\rho_\alpha$} and $d_\alpha$ are the quadrature weights and nodes for the particle size, respectively, $N_d$ is the number of size nodes, and $\delta$ denotes the Dirac delta function. For each size node $d_\alpha$, the conditional shape distribution is represented by
\begin{equation}
n_{\chi|d}(\chi|d_\alpha) \approx \sum_{\beta=1}^{N_\chi} \rho_{\beta|\alpha} \, \delta \!\left( \chi - \chi_{\beta|\alpha} \right),
\label{eq:shape_quadrature}
\end{equation}
where $\rho_{\beta|\alpha}$ and $ \chi_{\beta|\alpha}$ are the conditional quadrature weights and nodes for the particle shape, and $N_\chi$ is the number of shape nodes. The same procedure is applied recursively to the three velocity components, yielding
\begin{equation}
n \approx \sum_{\alpha=1}^{N_d} \rho_\alpha \delta(d-d_\alpha) \sum_{\beta=1}^{N_\chi} \rho_{\beta|\alpha} \delta(\chi-\chi_{\beta|\alpha}) \sum_{\gamma=1}^{N_w} \rho_{\gamma|\alpha\beta} \general{\delta(w-w_{\gamma|\alpha\beta})} \sum_{\mu=1}^{N_u} \rho_{\mu|\alpha\beta\gamma} \general{\delta(u-u_{\mu|\alpha\beta\gamma})} \sum_{\nu=1}^{N_v} \rho_{\nu|\alpha\beta\gamma\mu} \general{\delta(v-v_{\nu|\alpha\beta\gamma\mu})},
\label{eq:5d-cqmom}
\end{equation}
showing explicitly the dependence of each quadrature on the preceding conditioning variables. \general{In the following, the one-dimensional quadrature of a conditional distribution for a given combination of nodes of the preceding coordinates, for instance the quadrature of $u$ conditioned on $(d_\alpha, \chi_{\beta|\alpha}, w_{\gamma|\alpha\beta})$, is referred to as a branch of the conditional hierarchy.} Substituting Eq.~(\ref{eq:5d-cqmom}) into Eq.~(\ref{eq:5d-moment}) yields the moments
\begin{equation}
M_{ijklm} = \sum_{\alpha=1}^{N_d} \sum_{\beta=1}^{N_\chi} \sum_{\gamma=1}^{N_w} \sum_{\mu=1}^{N_u} \sum_{\nu=1}^{N_v} W_{\alpha\beta\gamma\mu\nu} \, d_\alpha^i \, \chi_{\beta|\alpha}^{\,j} \, \general{u_{\mu|\alpha\beta\gamma}^{\,k} \, v_{\nu|\alpha\beta\gamma\mu}^{\,l} \, w_{\gamma|\alpha\beta}^{\,m}},
\label{eq:moment_reconstruction}
\end{equation}
with
\begin{equation}
W_{\alpha\beta\gamma\mu\nu} = \rho_\alpha \rho_{\beta|\alpha} \rho_{\gamma|\alpha\beta} \rho_{\mu|\alpha\beta\gamma} \rho_{\nu|\alpha\beta\gamma\mu}.
\end{equation}

\subsection{Moment inversion}
\label{sec:cqmom_inversion}

The quadrature representation introduced in the previous section is fully defined by the quadrature nodes and weights. By exploiting the conditional structure of Eq.~(\ref{eq:conditional_ndf}), the inversion of the five-dimensional moment set is reduced to a hierarchy of one-dimensional inversions \citep{Yuan2011}. \reviewerone{In the present work, these one-dimensional inversions are performed with the adaptive extension of the algorithm of \citet{Wheeler1974} proposed by \citet{Yuan2011}, including the modifications described by \citet{Denner2026a}. The adaptive algorithm reduces the number of quadrature nodes of a one-dimensional quadrature if the moments do not support it: if a diagonal element of the Wheeler recursion is zero or negative within a tolerance, if the relative node separation or the ratio of the smallest to the largest weight falls below a threshold, or if a node lies outside a specified support. We use a threshold of $10^{-6}$ for both the node separation and the weight ratio, since a node with less than a millionth of the largest weight of its branch carries no meaningful information about the conditional distribution. In addition, the support of the particle size and aspect ratio is $d > 0$ and $0 < \chi \leq 1$, respectively. An accepted quadrature has positive weights and nodes within the defined support, and it reproduces the moments from which it was computed, which are therefore realizable on the support of the coordinate.}

The inversion procedure begins with the marginal size moments, $M_{i0000}$, from which the size nodes $d_\alpha$ and weights $\rho_\alpha$ are reconstructed. The reconstructed quadrature is required to satisfy
\begin{equation}
M_{i0000} = \sum_{\alpha=1}^{N_d} \rho_\alpha d_\alpha^i, \qquad i=0,\ldots,2N_d-1.
\end{equation}
Once the marginal size quadrature has been reconstructed using the adaptive Wheeler algorithm, the mixed moments $M_{ij000}$ are used to determine the conditional shape moments associated with each size node. The conditional shape moments are defined as
\begin{equation}
M_{j|\alpha} = \sum_{\beta=1}^{N_\chi} \rho_{\beta|\alpha} \chi_{\beta|\alpha}^{\,j},
\label{eq:conditional_shape_moments}
\end{equation}
such that the mixed moments may be expressed as
\begin{equation}
M_{ij000} = \sum_{\alpha=1}^{N_d} \rho_\alpha d_\alpha^i M_{j|\alpha}.
\label{eq:mixed_shape_moments}
\end{equation}
For each order $j$, the conditional moments $M_{j|\alpha}$ are obtained from the transported mixed moments by solving the Vandermonde system
\begin{equation}
\begin{bmatrix}
    1 & 1 & \cdots & 1 \\
    d_1 & d_2 & \cdots & d_{N_d}\\
    \vdots & \vdots & \ddots & \vdots \\
    d_1^{N_d-1} & d_2^{N_d-1} & \cdots & d_{N_d}^{N_d-1}
\end{bmatrix}
\begin{bmatrix}
    \rho_1 & 0 & \cdots & 0 \\
    0 & \rho_2 & \cdots & 0\\
    \vdots & \vdots & \ddots & \vdots \\
    0 & 0 & \cdots & \rho_{N_d}
\end{bmatrix}
\begin{bmatrix}
M_{j|1}\\
M_{j|2}\\
\vdots\\
M_{j|N_d}
\end{bmatrix}
=
\begin{bmatrix}
M_{0j000}\\
M_{1j000}\\
\vdots\\
M_{N_d-1,j,0,0,0}
\end{bmatrix}.
\label{eq:vandermonde}   
\end{equation}
\reviewerone{If the moments of a branch support only $n < N$ nodes, the conditional moments of its child branches are obtained from the square $n \times n$ Vandermonde system formed by the mixed moments of parent order below $n$. The Vandermonde systems are solved with the Bj\"orck-Pereyra algorithm \citep{Bjorck1970}.} The resulting conditional moment set is then inverted using the adaptive Wheeler algorithm to obtain the conditional shape nodes $\chi_{\beta|\alpha}$ and weights $\rho_{\beta|\alpha}$. The same procedure is subsequently applied to the conditional velocity distributions. Proceeding recursively through the conditional hierarchy, the conditional moments are inverted to obtain the quadrature nodes \general{$w_{\gamma|\alpha\beta}$, $u_{\mu|\alpha\beta\gamma}$, and $v_{\nu|\alpha\beta\gamma\mu}$}, as well as their corresponding conditional weights, yielding the complete five-dimensional conditional quadrature representation of the moments given in Eq.~\eqref{eq:moment_reconstruction}. \reviewerthree{The resulting quadrature reproduces the transported moments from which it is constructed, subject to realizability and numerical accuracy and, after an adaptive reduction of the node count, limited to the moments that the reduced quadrature supports. It remains an approximation of the NDF, from which unclosed higher-order and fractional moments can be approximated.} \reviewerone{The moments are recomputed from the accepted quadrature, so that the moments and the quadrature remain mutually consistent.}

\section{Solution algorithm}
\label{sec:solution}

Following the reconstruction of the quadrature, the transported moments must be advanced in time by solving the moment transport equations, Eq.~\eqref{eq:mte}. In principle, all phase-space contributions to the moment evolution can be integrated simultaneously within a single system of equations. In practice, however, quadrature-based moment models often involve a combination of continuous and discontinuous phase-space processes possessing fundamentally different physical and mathematical characteristics. For this reason, operator splitting has become the dominant approach for the numerical solution of moment transport equations \citep{Yuan2011, Yuan2012, Nguyen2016, Kong2017, Li2019a,Patel2019a,Heylmun2021, Li2022c}. By decomposing the moment evolution into physically distinct subproblems, numerical methods tailored to the characteristics of each operator can be employed. 
Despite their frequent adoption, the consequences of operator splitting on the evolution of the transported moments have received comparatively little attention. Continuous and discontinuous operators may act in competing directions in phase space and different coupling strategies can therefore produce different moment trajectories. This issue is particularly relevant for the non-spherical particle systems considered in the present work, where aerodynamic forces and attrition continuously modify the phase-space variables, while breakage simultaneously generates new particle populations and introduces kinetic energy through recoil. 

\reviewerthree{The different numerical treatment of the two classes of processes follows from the nature of their operators. Following Eq.~\eqref{eq:moment_reconstruction}, the NDF is approximated by $N_q = N_d N_\chi N_w N_u N_v$ quadrature nodes, denoted in the following by a single index $q$, with weights $W_q$ and abscissas $\boldsymbol{\xi}_q = (d_q, \chi_q, u_q, v_q, w_q)$. Integrating Eq.~\eqref{eq:continuous_processes} by parts, assuming vanishing boundary contributions, and applying the reconstructed quadrature, the contribution of the continuous processes becomes}
\begin{equation}
\reviewerthree{C_{ijklm} = \int_{\Omega} n \, \dot{\boldsymbol{\xi}} \cdot \nabla_{\boldsymbol{\xi}} \left( d^i \chi^j u^k v^l w^m \right) \,\text{d} \boldsymbol{\xi} \approx \sum_{q=1}^{N_q} W_q \, \dot{\boldsymbol{\xi}}(\boldsymbol{\xi}_q) \cdot \nabla_{\boldsymbol{\xi}} \left( d^i \chi^j u^k v^l w^m \right) \big|_{\boldsymbol{\xi}_q}.}
\label{eq:continuous_moment_source}
\end{equation}
\reviewerthree{This is exactly the rate of change of the moments of the quadrature if the weights are held constant and each node follows its trajectory in phase space,}
\begin{equation}
\reviewerthree{\frac{\text{d} \boldsymbol{\xi}_q}{\text{d} t} = \dot{\boldsymbol{\xi}} \left(\boldsymbol{\xi}_q \right).}
\label{eq:continuous_node}
\end{equation}
\reviewerthree{This advection in phase space, a local differential operator, therefore preserves the quadrature form of the NDF: each node evolves as a representative particle according to the particle model of Section~\ref{sec:model}. The velocity components are governed by Eq.~\eqref{eq:particle_acceleration} and the aerodynamic forces of Section~\ref{sec:forces}, whereas the diameter and aspect ratio are governed by the attrition model of Section~\ref{sec:attrition}. The moments are then recomputed from the updated quadrature.} 

\reviewerthree{The breakage operator, in contrast, is a nonlocal integral operator that removes parent particles and creates daughter particles at states that are not occupied by the existing nodes, so that its effect cannot be readily represented by moving the nodes. Its contribution is treated as a source term for the moments and a new quadrature is subsequently reconstructed.} Substituting the breakage source term defined in Eq.~\eqref{eq:breakage_source} into Eq.~\eqref{eq:discontinuous_processes} yields
\begin{equation}
D_{ijklm} = \int_{\Omega} R_{\text{break}}(\boldsymbol{\xi}) \,n(\boldsymbol{\xi}) \left[\bar{b}_{ijklm}(\boldsymbol{\xi}) - d^i \chi^j u^k v^l w^m \right] \,\text{d}\boldsymbol{\xi},
\label{eq:dmt_breakage_2}
\end{equation}
where $\bar{b}_{ijklm}$ denotes the moment of the daughter distribution. For the present breakage model, the size and shape of the daughter particles are defined by the model presented in Section~\ref{sec:breakage}, while the daughter velocity includes a stochastic recoil contribution. Assuming the recoil velocity to be isotropic and Gaussian distributed with standard deviation $\sigma_{\text{recoil}}$, the moments of the daughter velocity distribution can be evaluated analytically. For an arbitrary velocity component $u$, the $k$-th daughter moment is given by
\begin{equation}
\left\langle u_{\text{daughter}}^{\,k} \right\rangle = \sum_{p=0}^{\lfloor k/2 \rfloor} \binom{k}{2p} (2p-1)!! \, \sigma_{\text{recoil}}^{2p} \, (e \, u)^{\,k-2p},
\label{eq:recoil_moments}
\end{equation}
where $u$ is the corresponding velocity component of the parent particle. Analogous expressions are obtained for the velocity components $v$ and $w$. \general{Since the two daughter particles of a breakage event have the same size and shape and the recoil velocity components are independent, the moment of the daughter distribution $\beta(\cdot|\boldsymbol{\xi})$ of a parent particle in the state $\boldsymbol{\xi}$ follows from Eqs.~\eqref{eq:d_daughter}--\eqref{eq:chi_daughter_disk} and~\eqref{eq:recoil_moments} as}
\begin{equation}
\general{\bar{b}_{ijklm}(\boldsymbol{\xi}) = 2 \left[d_\text{daughter}(d)\right]^i \left[\chi_\text{daughter}(\chi)\right]^j \left\langle u_\text{daughter}^{\,k} \right\rangle \left\langle v_\text{daughter}^{\,l} \right\rangle \left\langle w_\text{daughter}^{\,m} \right\rangle,}
\label{eq:daughter_moments}
\end{equation}
\general{where the factor 2 accounts for the two daughter particles. The integral in Eq.~\eqref{eq:dmt_breakage_2} is evaluated with the quadrature reconstructed from the current moments:}
\begin{equation}
D_{ijklm} \approx \sum_{q=1}^{N_q} \general{W_q R_{\text{break}} (d_q)} \left[ \bar{b}_{ijklm}(\boldsymbol{\xi}_q) - d_q^i \chi_q^j u_q^k v_q^l w_q^m \right].
\label{eq:dmt_breakage_3}
\end{equation}

\general{Denoting the node advection by the operator $\mathcal{C}$ and the integration of the breakage source by the operator $\mathcal{D}$, the two first-order Lie splittings of a time step $\Delta t$ are
\begin{align}
\mathbf{M}^{n+1} &= \mathcal{D}_{\Delta t} \circ \mathcal{C}_{\Delta t} \left(\mathbf{M}^{n} \right),
\label{eq:operator-splitting-CD}
\\ 
\mathbf{M}^{n+1} &= \mathcal{C}_{\Delta t} \circ \mathcal{D}_{\Delta t} \left(\mathbf{M}^{n}\right), 
\label{eq:operator-splitting-DC}
\end{align}
as, for instance, applied by \citet{Li2022c} and \citet{Heylmun2021}, respectively,} where superscript $n$ is the time-step number. Since the two operators generally do not commute, the resulting solutions differ and both formulations are only first-order accurate in time \citep{Descombes2007}. To reduce the splitting error and recover second-order temporal accuracy, the present work employs a Strang splitting \citep{Strang1968},
\begin{equation}
\mathbf{M}^{n+1} = \mathcal{C}_{\Delta t/2} \circ \mathcal{D}_{\Delta t} \circ \mathcal{C}_{\Delta t/2} \left(\mathbf{M}^{n}\right),
\label{eq:strang}
\end{equation}
\general{following the work of \citet{Nguyen2016}}.

\reviewerone{Applying Strang splitting, a time step comprises four operations. First, the nodes of the current quadrature are advanced over half a time step according to Eq.~\eqref{eq:continuous_node}, with constant weights, and the moments $\mathbf{M}^{n+1/2}$ are recomputed from the advected nodes. The node trajectories are integrated with the embedded third-order Runge-Kutta (RK) scheme of \citet{Bogacki1989}, with adaptive substeps controlled by a relative tolerance of $10^{-6}$ and an absolute tolerance of $10^{-10}$. Second, the breakage source is integrated over the full time step with the third-order SSP-RK scheme of \citet{Shu1988}, as proposed by \citet{Nguyen2016}, in which the breakage contribution, Eq.~\eqref{eq:dmt_breakage_3}, is evaluated at each stage with the quadrature reconstructed from the stage moments. Hence, each of the three RK stages requires a CQMOM inversion. After every stage, the marginal moments of each coordinate are tested for realizability on its support: the half-line for $d$, the interval $[0,1]$ for $\chi$, and the real line for the velocity components \citep{Nguyen2016, Laurent2017, Fox2023a}. A failed test stops the simulation; the moments are neither clipped nor projected. Third, the quadrature is reconstructed from the moments at the end of the source integration and the moments are recomputed from it. Fourth, the nodes of this quadrature are advanced over the remaining half time step, and the moments $\mathbf{M}^{n+1}$ are recomputed from the advected nodes.}
\reviewertwo{Each time step thus requires four CQMOM inversions, three in the integration of the breakage source and one after it, whereas the node advection requires none. The same holds for the Lie splittings, Eqs.~\eqref{eq:operator-splitting-CD} and \eqref{eq:operator-splitting-DC}.} 

\newpage
\section{Results}
\label{sec:results}

\begin{table}[t]
    \centering
    \caption{\reviewertwo{Relative deviation (in $\%$) of the CQMOM solution, using the baseline quadrature and Strang splitting, from the mean of the Monte Carlo solution in the $L_2$ norm $\varepsilon_q^{(2)}$, Eq.~\eqref{eq:error_measure_2}, with the maximum norm $\varepsilon_q^{(\infty)}$, Eq.~\eqref{eq:error_measure_inf}, in parentheses. Considered are the number of particles $N_\mathrm{p}$, the mean values $\langle \cdot \rangle$ of each phase-space coordinate, the variance $\mathrm{Var}(\cdot)$ of the particle diameter and aspect ratio, as well as the kinetic energy $\mathbf{E}_\mathrm{k}$ and granular temperature $\boldsymbol{\Theta}$ of the particles.}}
    \label{tab:errors}
    \reviewertwo{\begin{tabular}{l cc cc cc}
    \hline
    & \multicolumn{2}{c}{Test case 1} & \multicolumn{2}{c}{Test case 2} & \multicolumn{2}{c}{Test case 3} \\
    $q$ & $\text{St} = 0.1$ & $\text{St} = 10$ & $\text{St} = 0.1$ & $\text{St} = 10$ & $\text{St} = 0.1$ & $\text{St} = 10$ \\
    \hline
    $N_\text{p}$ & 0.19 (0.21) & 0.24 (0.29) & 0.081 (0.10) & 0.19 (0.20) & 0.18 (0.22) & 0.28 (0.34) \\
    $\langle d \rangle$ & 0.086 (0.12) & 0.094 (0.11) & 0.11 (0.15) & 0.095 (0.13) & 0.086 (0.13) & 0.10 (0.12) \\
    $\langle \chi \rangle$ & 0.15 (0.24) & 0.18 (0.26) & 0.091 (0.24) & 0.14 (0.24) & 0.073 (0.16) & 0.12 (0.20) \\
    $\langle u \rangle$ & 0.096 (0.12) & 0.78 (0.98) & 0.095 (0.17) & 0.53 (0.56) & 0.075 (0.18) & 0.30 (0.32) \\
    $\langle v \rangle$ & 0.061 (0.10) & 0.61 (0.87) & 0.11 (0.33) & 1.3 (1.4) & 0.093 (0.25) & 0.59 (0.51) \\
    $\langle w \rangle$ & 0.26 (0.62) & 0.43 (0.80) & 0.38 (1.3) & 0.42 (0.79) & 0.43 (1.4) & 0.20 (0.31) \\
    $\mathrm{Var}(d)$ & 1.1 (1.5) & 0.62 (0.82) & 1.2 (1.5) & 1.1 (1.3) & 1.2 (1.4) & 0.47 (1.0) \\
    $\mathrm{Var}(\chi)$ & 5.1 (10) & 26 (29) & 5.8 (11) & 26 (28) & 0.22 (0.58) & 5.1 (7.9) \\
    $E_{\text{k},x}$ & 0.46 (0.64) & 0.45 (0.51) & 0.43 (0.70) & 0.38 (0.43) & 0.39 (0.67) & 0.44 (0.47) \\
    $E_{\text{k},y}$ & 0.35 (0.44) & 0.37 (0.31) & 0.53 (0.96) & 1.5 (1.2) & 0.47 (1.1) & 1.8 (1.6) \\
    $E_{\text{k},z}$ & 1.7 (1.5) & 1.2 (2.0) & 1.7 (2.4) & 1.1 (2.0) & 0.71 (0.57) & 0.18 (0.25) \\
    $\Theta_x$ & 2.9 (0.99) & 1.0 (0.63) & 2.0 (0.70) & 1.0 (1.2) & 4.0 (1.0) & 0.91 (1.1) \\
    $\Theta_y$ & 3.9 (2.7) & 1.8 (1.3) & 3.6 (2.0) & 1.5 (1.3) & 3.0 (1.9) & 1.5 (1.6) \\
    $\Theta_z$ & 2.9 (0.61) & 6.9 (7.0) & 1.9 (0.57) & 4.0 (6.8) & 4.3 (0.97) & 0.93 (1.1) \\
    \hline
    \end{tabular}}
\end{table}

\begin{figure}[t]
    \centering
    \begin{subfigure}{\linewidth}
        \includegraphics[width=\linewidth]{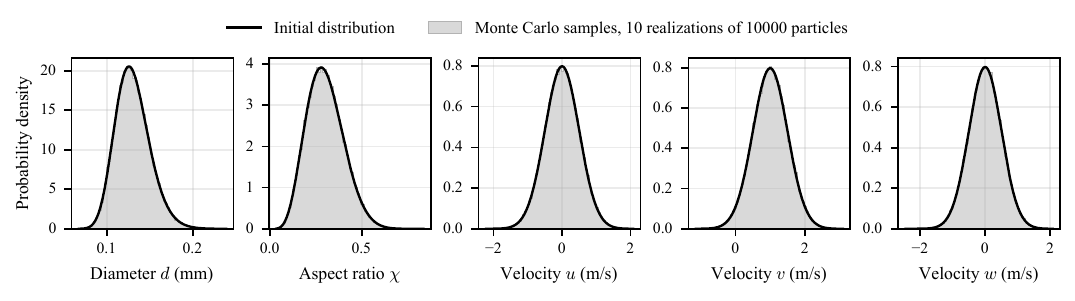}
        \caption{Initial population for $\text{St}=0.1$.}
    \end{subfigure}\\[0.5em]
    \begin{subfigure}{\linewidth}
        \includegraphics[width=\linewidth]{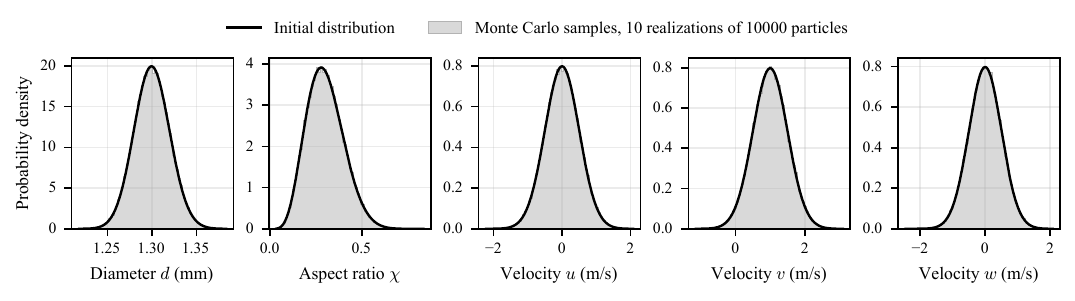}
        \caption{Initial population for $\text{St}=10$.}
    \end{subfigure}
    \caption{Initial populations for particle settling and breakage in a flow\general{: marginal densities of the initial distribution, from whose exact moments the CQMOM simulations start, and histograms of the pooled initial samples of the ten Monte Carlo realizations}.}
    \label{fig:test1_initial}
\end{figure}

\reviewerone{All simulations of this work are performed with the open-source C++ library QMOMKit \citep{Denner2026a}, version v1.0.0\footnote{https://github.com/polycfd/QMOMKit/tree/v1.0.0, https://doi.org/10.5281/zenodo.23728106}, which implements the algorithm described above. The case-specific code, the configuration files, and the scripts required to reproduce all results, figures, and tables of this work are available in a separate repository, given in the data availability statement.}
To systematically assess the ability of the proposed CQMOM framework to represent coupled geometric and kinematic dynamics of non-spherical particle populations, three test cases of increasing complexity are considered. The underlying particle model \reviewerthree{for oblate spheroids} combines continuous processes (drag and attrition) with discontinuous processes (particle breakage and recoil) acting on the geometric ($d,\chi$) and kinematic ($u,v,w$) dimensions of the considered phase space. The first test case (Section \ref{sec:test1}) models particle settling and breakage in a flow, thereby combining continuous kinematic evolution due to aerodynamic forces with discontinuous geometric evolution due to breakage. The second test case (Section \ref{sec:test2}) further introduces crossing particle jets as well as a discontinuous kinematic evolution through breakage-induced recoil. The final test case (Section \ref{sec:test3}) also includes continuous attrition of the particles, resulting in fully coupled evolution of particle size, shape, and velocity within the five-dimensional phase space. \general{The influence of the chosen operator splitting, the time step, the quadrature resolution, and the conditioning order, as well as the computational cost, is discussed in more detail in Section~\ref{sec:discussion}.}

The carrier fluid is assumed to be air, with a density of $\rho_\text{f} = 1.2 \ \text{kg m}^{-3}$ and a dynamic viscosity of $\mu_\text{f} = 1.8 \times 10^{-5} \ \text{Pa s}$. It rotates with a time-dependent velocity of $\mathbf{u}_\text{f} = ( U \cos(2 \pi f  t), U \sin(2 \pi f t), 0)$, with amplitude $U = 2 \ \text{m s}^{-1}$ and oscillation frequency $f = 1 \ \text{s}^{-1}$, and the gravitational acceleration is $\mathbf{g} = (0 , 0, -9.81 ) \ \text{m s}^{-2}$. The particles have a material density of $\rho_\text{p} = 2000 \ \text{kg m}^{-3}$ and a volume-equivalent mean diameter of $\langle d \rangle \in \{1.3 \times 10^{-4}, 1.3 \times 10^{-3} \} \ \text{m}$, such that the Stokes number is $\text{St} = \langle d \rangle^2 \rho_\text{p} f/(18 \mu_\text{f}) \in \{0.1,10\}$. The breakage rate in Eq.~\eqref{eq:Rbreak} is $k_\text{break} = 10^8 \, \text{s}^{-1} \ \text{m}^{-2}$ for $\langle d \rangle = 1.3 \times 10^{-4} \ \text{m}$ and \general{$k_\text{break} = 10^6 \, \text{s}^{-1} \ \text{m}^{-2}$} for $\langle d \rangle = 1.3 \times 10^{-3} \ \text{m}$\general{, such that both populations have the same breakage number of the mean diameter, $k_\text{break} \langle d \rangle^2 / f = 1.69$}.

\reviewertwo{The baseline quadrature employs two nodes in each coordinate, $(N_d, N_\chi, N_u, N_v, N_w) = (2,2,2,2,2)$, i.e., $N_q = 32$ nodes,} \reviewerone{with the conditioning order $(d, \chi, w, u, v)$ of Section~\ref{sec:cqmom}.} 
\general{All CQMOM simulations start from the exact moments of the initial distribution and use the Strang splitting of Section~\ref{sec:solution} with a time step of $\Delta t = 5 \ \text{ms}$ up to $t = 1 \ \text{s}$.} \reviewertwo{This time step is a compromise rather than a converged choice: larger time steps increase the splitting error, whereas smaller time steps amplify individual outlier nodes of the velocity quadrature, as discussed together with the quadrature resolution and the conditioning order in Section~\ref{sec:discussion}.}

As a reference solution, a Monte Carlo method\footnote{The Monte Carlo reference solutions are produced with the Monte Carlo companion code of QMOMKit \citep{Denner2026a}.} is employed\general{, which is independent of the moment method}. The Monte Carlo method tracks each particle in the five-dimensional phase space ($d,\chi,u,v,w$) according to the particle model described in Section~\ref{sec:model}\general{: the trajectory of each particle is integrated with the same Runge-Kutta scheme and tolerances as the quadrature nodes, described in Section~\ref{sec:solution}, up to its next breakage event, at which two daughter particles replace it.} Since no moment closure or quadrature reconstruction is required, the Monte Carlo solution provides an independent representation of the underlying population balance and serves as the reference against which the presented CQMOM formulation is evaluated. 
\reviewertwo{The Monte Carlo reference solution is the mean of ten independent realizations, each starting from $10^4$ particles sampled from the initial distribution. For every considered quantity $q$, the deviation of the CQMOM solution from the Monte Carlo mean $\bar{q}_\text{mc}$ is measured over the output times $t_i$, at intervals of $10 \ \text{ms}$, in the $L_2$ norm and in the maximum norm,
\begin{align}
\varepsilon_q^{(2)} &= \left( \frac{\sum_i \left[ q_\text{cqmom}(t_i) - \bar{q}_\text{mc}(t_i) \right]^2}{\sum_i \bar{q}_\text{mc}(t_i)^2} \right)^{1/2}, \label{eq:error_measure_2} 
\\
\varepsilon_q^{(\infty)} &= \frac{\max_i \left| q_\text{cqmom}(t_i) - \bar{q}_\text{mc}(t_i) \right|}{\max_i \left| \bar{q}_\text{mc}(t_i) \right|}.
\label{eq:error_measure_inf}
\end{align}
Table~\ref{tab:errors} lists both measures for all test cases and quantities considered below. The statistical uncertainty of the reference, defined as the half-width of the 95\,\% confidence interval of the Monte Carlo mean obtained from the corresponding Student's $t$ distribution \citep{Student1908}, amounts in the $L_2$ norm to at most 0.4\,\% for the particle number, the mean diameter, and the mean aspect ratio, and to at most 1.6\,\% for all other quantities.}

\subsection{Particle settling and breakage in a flow}
\label{sec:test1}

\begin{figure}[t]
    \centering
    \begin{subfigure}{\linewidth}
        \includegraphics[width=\linewidth]{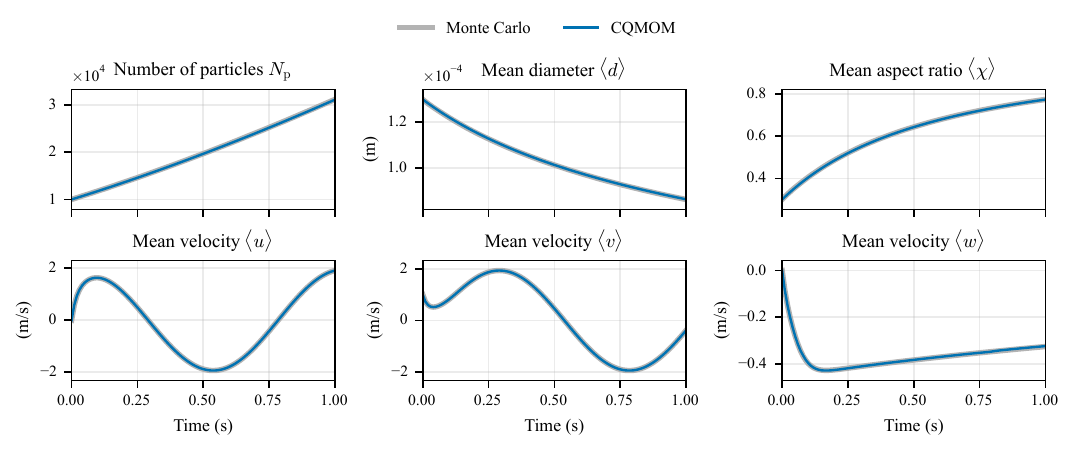}
        \caption{Evolution of the particle number, mean particle size, aspect ratio, and velocity components.} 
        \label{fig:test1_st01_moments}
    \end{subfigure}\\[0.5em]
    \begin{subfigure}{0.96\linewidth}
        \includegraphics[width=\linewidth]{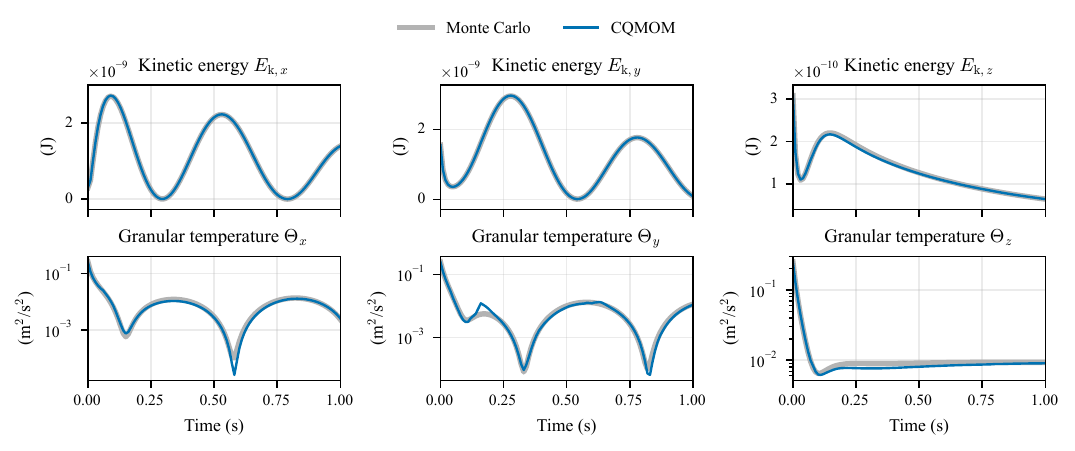}
        \caption{Evolution of the directional kinetic energies and granular temperatures of the particle population.}
        \label{fig:test1_st01_physics}
    \end{subfigure}
    \caption{Comparison of the Monte Carlo reference solution\general{ (the mean of ten realizations) and the} CQMOM for particle settling and breakage in a flow with $\text{St}=0.1$.}
    \label{fig:test1_st01}
\end{figure}

\begin{figure}[t]
    \centering
    \begin{subfigure}{\linewidth}
        \includegraphics[width=\linewidth]{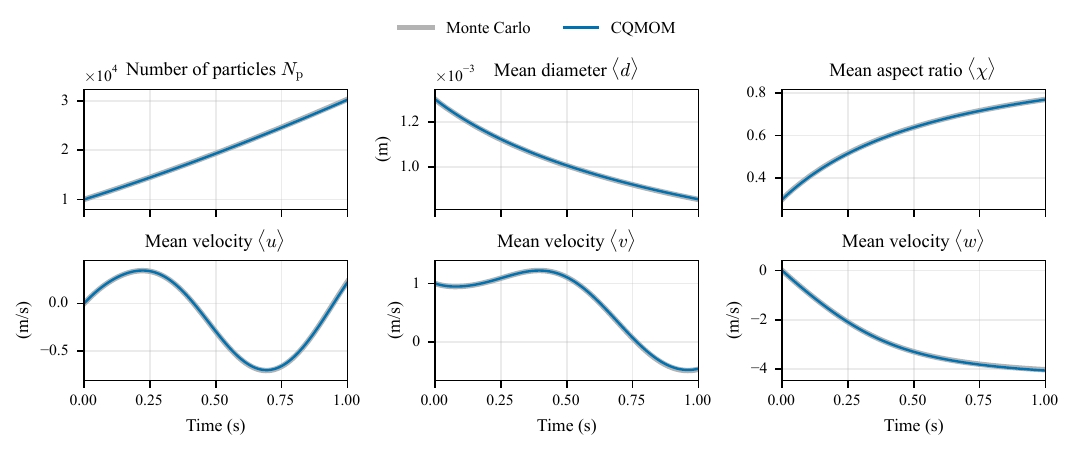}
        \caption{Evolution of the particle number, mean particle size, aspect ratio, and velocity components.} 
        \label{fig:test1_st10_moments}
    \end{subfigure}\\[0.5em]
    \begin{subfigure}{0.96\linewidth}
        \includegraphics[width=\linewidth]{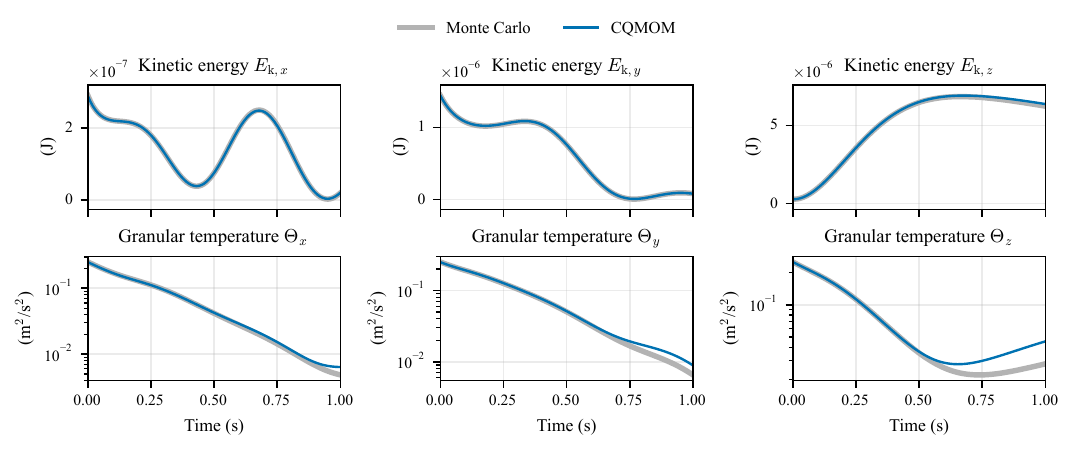}
        \caption{Evolution of the directional kinetic energies and granular temperatures of the particle population.}
        \label{fig:test1_st10_physics}
    \end{subfigure}
    \caption{Comparison of the Monte Carlo reference solution\general{ (the mean of ten realizations) and the} CQMOM for particle settling and breakage in a flow with $\text{St}=10$.}
    \label{fig:test1_st10}
\end{figure}

The first test case considers a population of oblate spheroids evolving in a time-dependent rotating fluid velocity field under the combined action of drag, gravity, and binary breakage, thereby isolating the interaction between continuous kinematic evolution and discontinuous geometric evolution. Between breakage events, the particle velocity evolves continuously according to drag and gravitational acceleration, following Eq.~\eqref{eq:particle_acceleration}, while the particle geometry remains unchanged ($\text{d}d/\text{d}t = 0$, $\text{d}\chi/\text{d}t = 0$). Simultaneously, binary breakage introduces daughter particles with a smaller size, following Eq.~\eqref{eq:d_daughter}, and a larger aspect ratio (i.e., more spherical), following  Eq.~\eqref{eq:chi_daughter_spherical} with $\alpha =0.5$, while the daughter particles inherit the velocity of their parent ($e = 1$, $\mathbf{v}_\text{recoil} = \mathbf{0}$). \general{The initial distribution, shown in Figure \ref{fig:test1_initial}, is the product of independent marginal distributions.} The particle diameters \general{follow} a log-normal distribution centered around $\langle d \rangle \in \{1.3 \times 10^{-4}, 1.3 \times 10^{-3} \} \ \text{m}$, with a standard deviation of $2 \times 10^{-5} \ \text{m}$, whereas the particle aspect ratios follow a beta distribution with a mean of $\langle \chi \rangle =0.3$ and a standard deviation of $0.1$. The particle velocities are initialized from independent normal distributions with a mean of $( \langle u \rangle,\langle v \rangle,\langle w \rangle ) = (0, 1, 0) \ \text{m s}^{-1}$ and a standard deviation of $0.5 \, \text{m s}^{-1}$ in each direction.

Figures \ref{fig:test1_st01} and \ref{fig:test1_st10} show, for particle populations with Stokes numbers of $\text{St}=0.1$ and $\text{St}=10$, respectively, the evolution of the zeroth-order moment $N_\text{p}=M_{00000}$, the first-order moments of the phase-space variables, as well as the particle kinetic energy $\mathbf{E}_{\text{k}}$ and granular temperature $\boldsymbol{\Theta}$. The particle kinetic energy and granular temperature in the $x$-direction are defined as
\begin{align}
    E_{\text{k},x} &= \general{\left\langle \frac{1}{2} \rho_\text{p} V_\text{p} u^2 \right\rangle} = \frac{\pi \rho_\text{p}}{12} \, \frac{M_{30200}}{M_{00000}}\\
    \Theta_x &= \langle u^2 \rangle - \langle u \rangle^2 = \frac{M_{00200}}{M_{00000}} - \left(\frac{M_{00100}}{M_{00000}}\right)^2,
\end{align}
respectively, and similarly for the physical-space coordinates $y$ and $z$. \general{For CQMOM, the moments $M_{30200}$, $M_{30020}$, and $M_{30002}$ are not among the transported moments and are, thus, evaluated from the quadrature, so that the kinetic energy also tests the quadrature closure.} Together, these quantities provide a comprehensive assessment of the ability of the proposed framework to capture both the bulk evolution of the particle population and the underlying phase-space correlations. The zeroth-order moment verifies the conservation and evolution of the number of particles, while the first-order moments characterize the mean evolution of particle size, shape, and velocity. The particle kinetic energy probes the preservation of coupled size-velocity correlations through its simultaneous dependence on particle mass and velocity, whereas the granular temperature measures the velocity variance of the population and is, thus, sensitive to the generation of velocity fluctuations. \reviewertwo{In addition, Table~\ref{tab:errors} lists the variances of the diameter and of the aspect ratio, 
\begin{align}
    \mathrm{Var}(d) &= \frac{M_{20000}}{M_{00000}} - \left(\frac{M_{10000}}{M_{00000}}\right)^2,\\
    \mathrm{Var}(\chi) &= \frac{M_{02000}}{M_{00000}} - \left(\frac{M_{01000}}{M_{00000}}\right)^2.
\end{align}
In particular, $\mathrm{Var}(\chi)$ tests the representation of the conditional aspect-ratio distribution that is further discussed in  Section~\ref{sec:discussion_resolution}.} 

\reviewertwo{For both Stokes numbers, the mean values of the phase space coordinates are qualitatively in good agreement, see Figures \ref{fig:test1_st01_moments} and \ref{fig:test1_st10_moments}. The particle number, the mean diameter, and the mean aspect ratio deviate from the Monte Carlo mean by at most 0.24\,\% in the $L_2$ norm, see Table~\ref{tab:errors}, which is within the statistical uncertainty of the reference. The mean velocities deviate by at most 0.8\,\% and the kinetic energies by at most 1.7\,\%. The granular temperatures are the most sensitive quantities, with deviations of 2.9--3.9\,\% for $\text{St}=0.1$ and up to 6.9\,\% for $\text{St}=10$. For $\text{St}=0.1$, $\Theta_y$ exhibits a temporary overshoot at $t \approx 0.16 \ \text{s}$ in Figure~\ref{fig:test1_st01_physics}, whereas for $\text{St}=10$, the granular temperatures, in particular $\Theta_z$, are overpredicted at late times, see Figure~\ref{fig:test1_st10_physics}, when the velocity variance has decayed by an order of magnitude. The variance of the aspect ratio deviates by 5.1\,\% for $\text{St}=0.1$ and by 26\,\% for $\text{St}=10$, as reported in Table \ref{tab:errors}. The origins of these deviations are discussed in Section~\ref{sec:discussion}.}

\subsection{Multimodal particle jet with stochastic breakup recoil}
\label{sec:test2}

\begin{figure}[t]
    \centering
    \begin{subfigure}{\linewidth}
        \includegraphics[width=\linewidth]{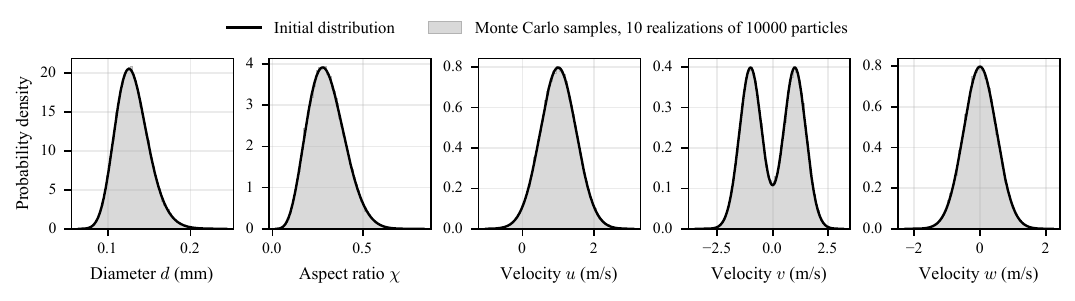}
        \caption{Initial population for $\text{St}=0.1$.}
    \end{subfigure}\\[0.5em]
    \begin{subfigure}{\linewidth}
        \includegraphics[width=\linewidth]{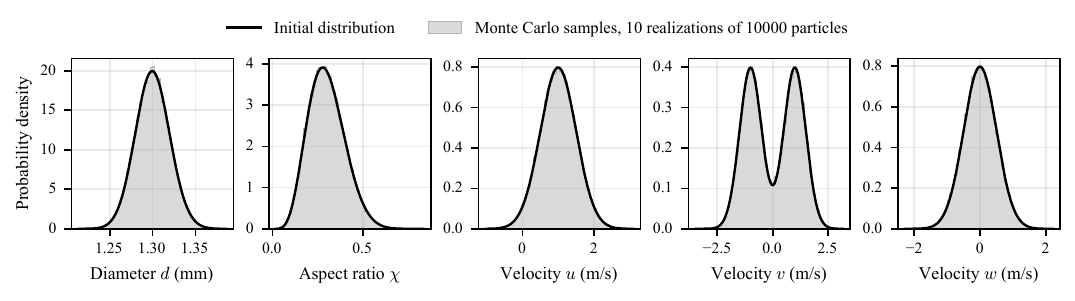}
        \caption{Initial population for $\text{St}=10$.}
    \end{subfigure}
    \caption{Initial populations for the multimodal particle jet with stochastic breakup recoil\general{ and for particle attrition and breakage in a flow: marginal densities of the initial distribution, from whose exact moments the CQMOM simulations start, and histograms of the pooled initial samples of the ten Monte Carlo realizations}.}
    \label{fig:test2_initial}
\end{figure}

\begin{figure}[t]
    \centering
    \begin{subfigure}{\linewidth}
        \includegraphics[width=\linewidth]{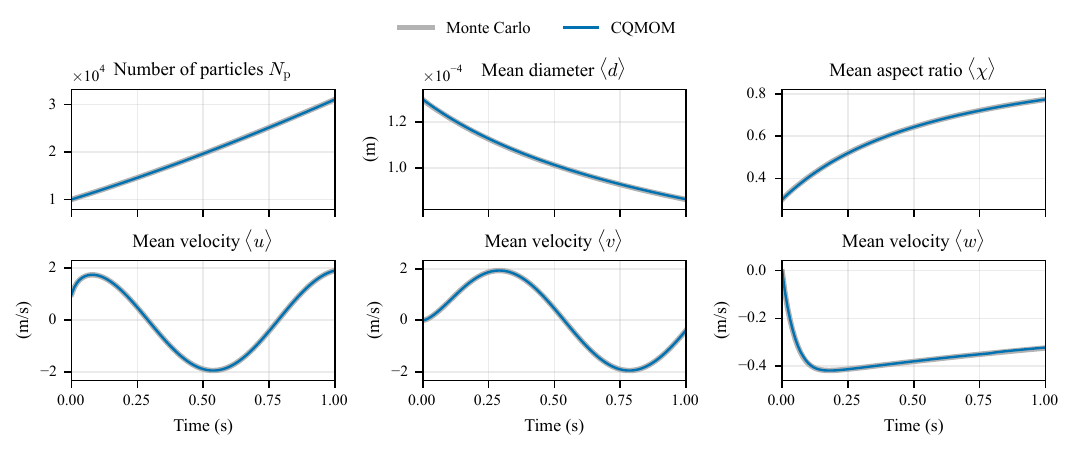}
        \caption{Evolution of the particle number, mean particle size, aspect ratio, and velocity components.} 
        \label{fig:test2_st01_moments}
    \end{subfigure}\\[0.5em]
    \begin{subfigure}{0.96\linewidth}
        \includegraphics[width=\linewidth]{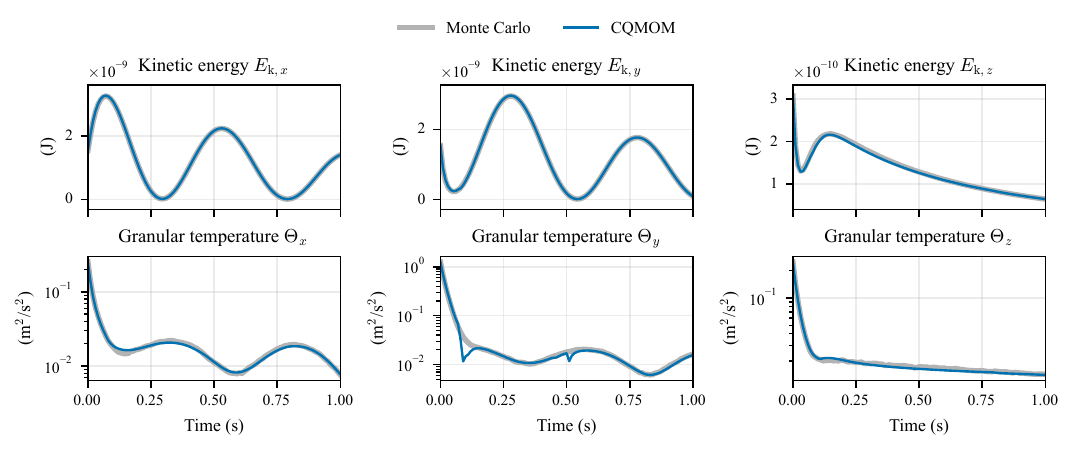}
        \caption{Evolution of the directional kinetic energies and granular temperatures of the particle population.}
        \label{fig:test2_st01_physics}
    \end{subfigure}
    \caption{Comparison of the Monte Carlo reference solution\general{ (the mean of ten realizations) and the} CQMOM for the multimodal particle jet with stochastic breakup recoil with $\text{St}=0.1$.}
    \label{fig:test2_st01}
\end{figure}

\begin{figure}[t]
    \centering
    \begin{subfigure}{\linewidth}
        \includegraphics[width=\linewidth]{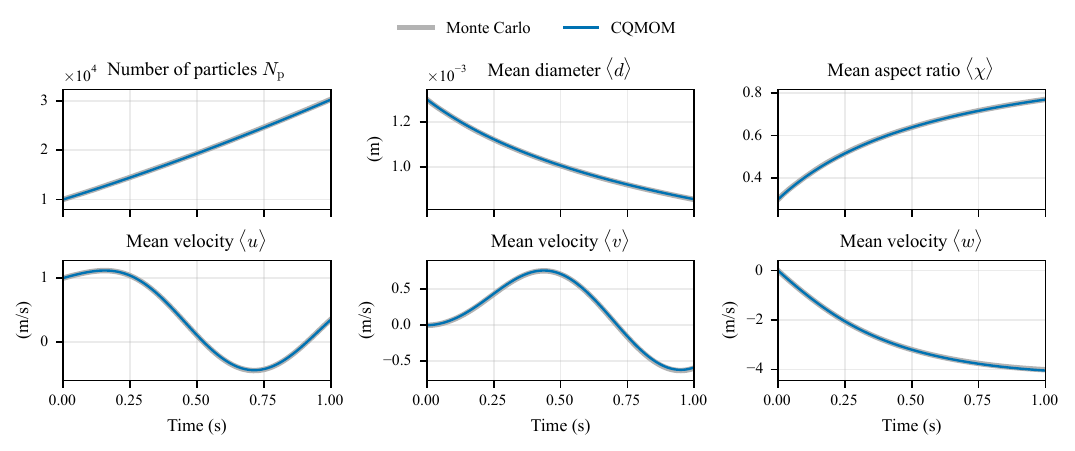}
        \caption{Evolution of the particle number, mean particle size, aspect ratio, and velocity components.} 
        \label{fig:test2_st10_moments}
    \end{subfigure}\\[0.5em]
    \begin{subfigure}{\linewidth}
        \includegraphics[width=\linewidth]{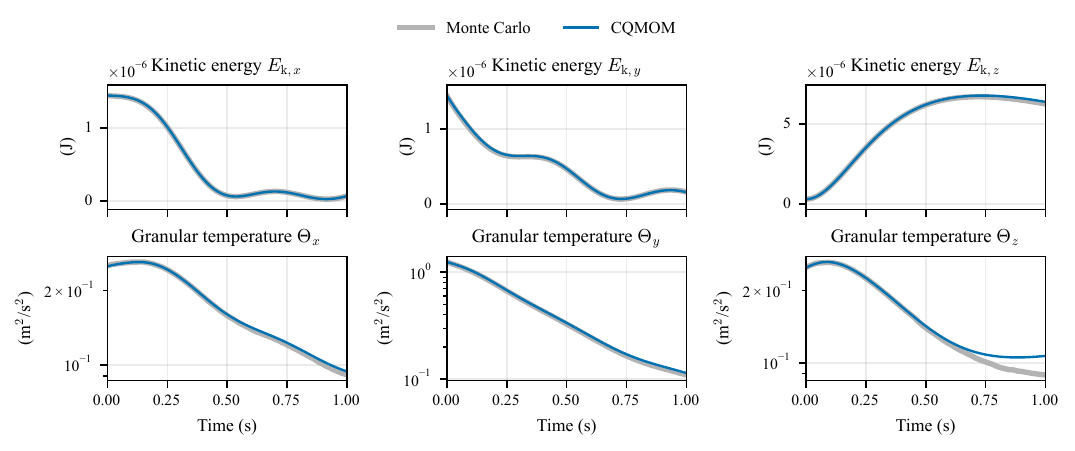}
        \caption{Evolution of the directional kinetic energies and granular temperatures of the particle population.}
        \label{fig:test2_st10_physics}
    \end{subfigure}
    \caption{Comparison of the Monte Carlo reference solution\general{ (the mean of ten realizations) and the} CQMOM for the multimodal particle jet with stochastic breakup recoil with $\text{St}=10$.}
    \label{fig:test2_st10}
\end{figure}

Compared to the first test case, the second test case introduces two additional sources of complexity. First, the initial particle population, shown in Figure \ref{fig:test2_initial}, consists of two counter-propagating jets with a bimodal distribution of the $v$-velocity, consisting of two independent normal distributions centered at $-1 \ \text{m s}^{-1}$ and $+1 \ \text{m s}^{-1}$, both with a standard deviation of $0.5 \ \text{m s}^{-1}$. The initial distribution of the $u$-velocity is now centered on $1 \ \text{m s}^{-1}$, also with a standard deviation of $0.5 \ \text{m s}^{-1}$. Second, each breakage event generates a stochastic recoil that continuously injects kinetic energy and velocity variance into the particle population. This recoil is modelled through the recoil model given in Eq.~\eqref{eq:v_recoil}, with \general{$e = 1$} and $\sigma_\text{recoil} = 0.5 \ \text{m s}^{-1}$, whereby the daughter velocity moments are defined by Eq.~\eqref{eq:recoil_moments}. Together, these mechanisms create a substantially more demanding test case, requiring the moment closure to accurately represent both multimodal velocity distributions and the generation of higher-order velocity moments.

Figures \ref{fig:test2_st01} and \ref{fig:test2_st10} show the results for the two particle populations with Stokes numbers of $\text{St}=0.1$ and $\text{St}=10$, respectively. The influence of the additional mechanisms is most apparent in the evolution of the kinetic energies and granular temperatures, as the present configuration directly probes the ability of the moment algorithm to represent the velocity dispersion generated by the stochastic recoil. \reviewertwo{As in the first test case, the particle number, the mean diameter, and the mean aspect ratio deviate from the Monte Carlo mean by at most 0.2\,\%. Despite the velocity dispersion generated by the recoil, the mean velocities deviate by at most 1.3\,\%, the kinetic energies by at most 1.7\,\%, and the granular temperatures by 1.0--3.6\,\%, and by 4.0\,\% for $\Theta_z$ for $\text{St}=10$, which is again overpredicted at late times, see Figure~\ref{fig:test2_st10_physics}. For $\text{St}=0.1$, $\Theta_y$ exhibits brief dips at $t \approx 0.09 \ \text{s}$ and $t \approx 0.51 \ \text{s}$, as observed in Figure~\ref{fig:test2_st01_physics}. The variance of the aspect ratio deviates by 5.8\,\% and 26\,\%, similar to the first test case.}

\subsection{Particle attrition and breakage in a flow}
\label{sec:test3}

\begin{figure}[t]
    \centering
    \begin{subfigure}{\linewidth}
        \includegraphics[width=\linewidth]{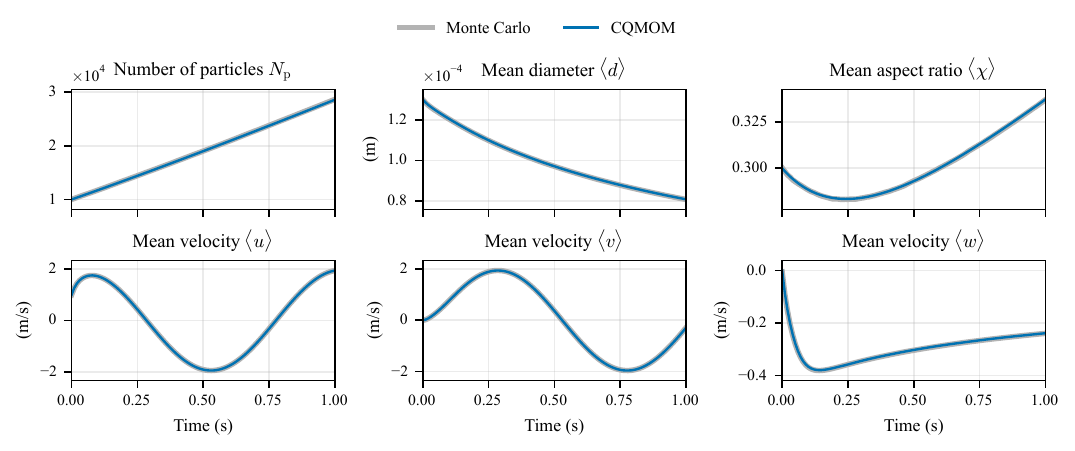}
        \caption{Evolution of the particle number, mean particle size, aspect ratio, and velocity components.} 
        \label{fig:test3_st01_moments}
    \end{subfigure}\\[0.5em]
    \begin{subfigure}{0.96\linewidth}
        \includegraphics[width=\linewidth]{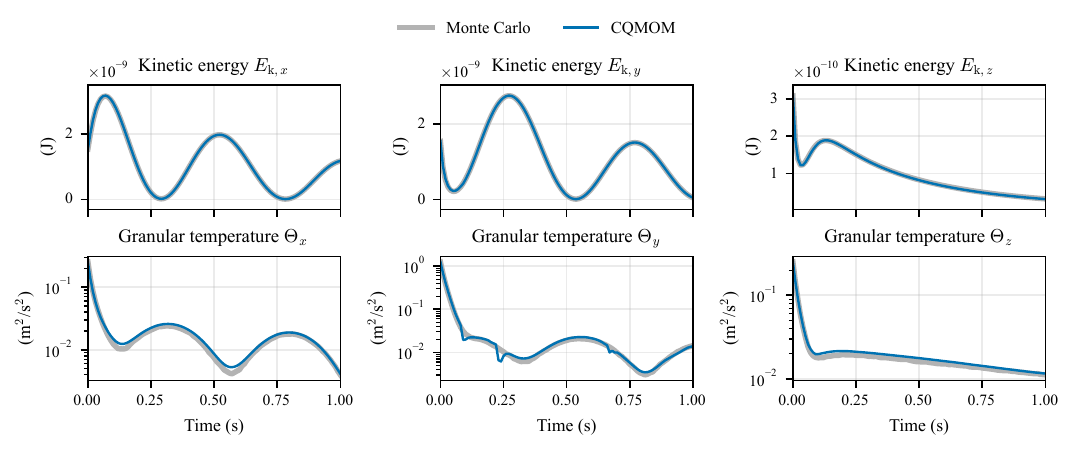}
        \caption{Evolution of the directional kinetic energies and granular temperatures of the particle population.}
        \label{fig:test3_st01_physics}
    \end{subfigure}
    \caption{Comparison of the Monte Carlo reference solution\general{ (the mean of ten realizations) and the} CQMOM for particle attrition and breakage in a flow with $\text{St}=0.1$.}
    \label{fig:test3_st01}
\end{figure}

\begin{figure}[t]
    \centering
    \begin{subfigure}{\linewidth}
        \includegraphics[width=\linewidth]{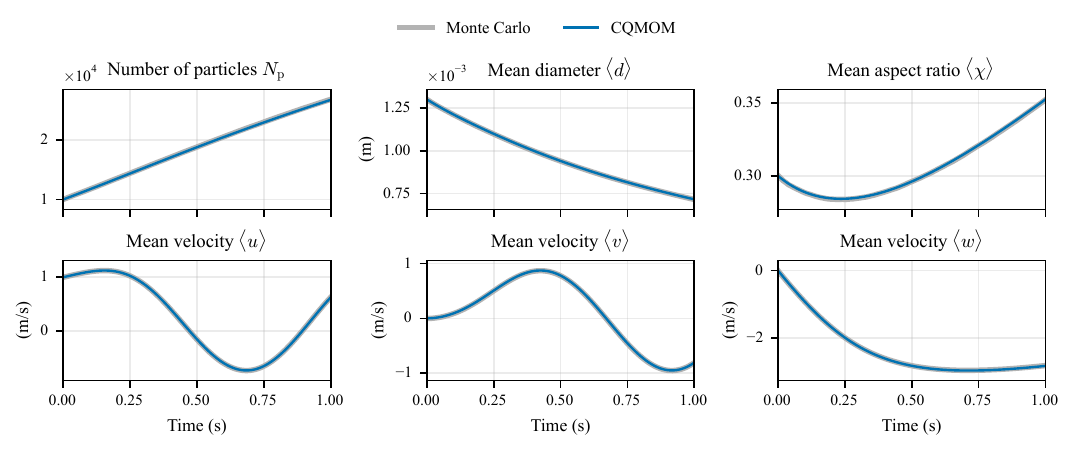}
        \caption{Evolution of the particle number, mean particle size, aspect ratio, and velocity components.} 
        \label{fig:test3_st10_moments}
    \end{subfigure}\\[0.5em]
    \begin{subfigure}{\linewidth}
        \includegraphics[width=\linewidth]{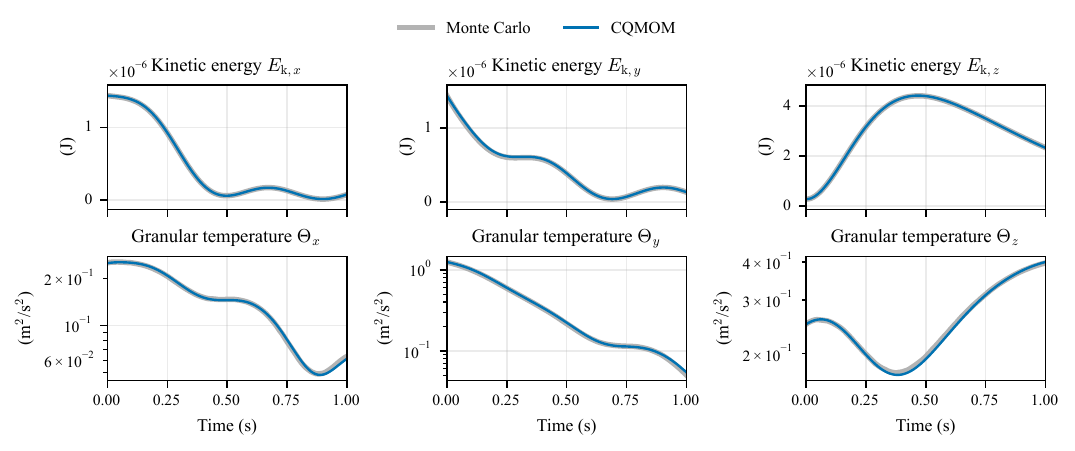}
        \caption{Evolution of the directional kinetic energies and granular temperatures of the particle population.}
        \label{fig:test3_st10_physics}
    \end{subfigure}
    \caption{Comparison of the Monte Carlo reference solution\general{ (the mean of ten realizations) and the} CQMOM for particle attrition and breakage in a flow with $\text{St}=10$.}
    \label{fig:test3_st10}
\end{figure}

The third test case combines all \general{model components} introduced in Section~\ref{sec:model}. In addition to aerodynamic transport, binary breakage, and stochastic  recoil, particles now undergo continuous surface attrition. This attrition is modelled by Eqs.~\eqref{eq:d_continuous} and \eqref{eq:chi_continuous}, with \reviewerthree{attrition coefficient $k_\text{attrition} = 3 \times 10^{-5} \ \text{s m}^{-1}$, $U_s = 5 \ \text{m s}^{-1}$, $d_\text{min} = 0$, and $d_s = 10^{-5} \ \text{m}$,} and shape relaxation timescale $\tau_\text{shape} = 2 \ \text{s}$. \reviewerthree{Compared with the second test case, attrition reduces the mean diameter at $t = 1 \ \text{s}$ by about 6\,\% at $\text{St}=0.1$ and by about 16\,\% at $\text{St}=10$. Since the particles remain far larger than $d_s$, the taper toward $d_\text{min}$ only ensures that the diameter remains positive.} The initial particle population is initialized with the same distributions as employed in Section \ref{sec:test2}, with initial distributions equivalent to those shown in Figure \ref{fig:test2_initial}. In this test case, however, breakage generates progressively flatter particles by reducing the aspect ratio following Eq.~\eqref{eq:chi_daughter_disk} with $\alpha = 0.5$, whereas the shape relaxation included in the attrition model drives particles toward a spherical geometry following Eq.~\eqref{eq:chi_continuous}. Consequently, \general{the present test case is the most demanding of the three:} all processes act simultaneously on both the geometric and kinematic dimensions of the phase space and, in the case of the aspect ratio, operate in opposing directions.

\reviewertwo{Figures~\ref{fig:test3_st01} and \ref{fig:test3_st10} show the results for both Stokes numbers. Despite the increased complexity of this test case, the results for all considered quantities are in very good qualitative agreement with the Monte Carlo reference solution. The deviations from the Monte Carlo solution listed in Table~\ref{tab:errors} are of the same magnitude as in the previous test cases: at most 0.28\,\% for the particle number, the mean diameter, and the mean aspect ratio, 0.6\,\% for the mean velocities, 1.8\,\% for the kinetic energies, and 4.3\,\% for $\text{St}=0.1$ and 1.5\,\% for $\text{St}=10$ for the granular temperatures. In contrast to the first two test cases, $\Theta_z$ for $\text{St}=10$ agrees with the Monte Carlo solution to within 0.9\,\%, and the variance of the aspect ratio deviates by only 0.2\,\% for $\text{St}=0.1$ and 5.1\,\% for $\text{St}=10$.}

\section{Discussion}
\label{sec:discussion}

\general{With the baseline quadrature (2 nodes per coordinate), Strang splitting, and a time step of 5~ms, the CQMOM solution is able to reproduce the particle number, the mean diameter, and the mean aspect ratio of the Monte Carlo solution within its statistical uncertainty, and the velocity statistics to within 7\,\%. The largest deviations are observed in the granular temperatures and in the variance of the aspect ratio. This section examines the dependence of these results on the operator splitting and the time step (Section~\ref{sec:discussion_splitting}), as well as on the quadrature resolution and the conditioning order (Section~\ref{sec:discussion_resolution}), and discusses the realizability of the moments and conditioning of the quadrature reconstruction (Section~\ref{sec:discussion_realizability}), the computational cost (Section~\ref{sec:discussion_cost}), as well as the scope of the particle model and extensions of the method (Section~\ref{sec:discussion_scope}).}

\subsection{Operator splitting and time step}
\label{sec:discussion_splitting}

\begin{figure}[t]
    \centering
    \begin{subfigure}{\linewidth}
        \includegraphics[width=\linewidth]{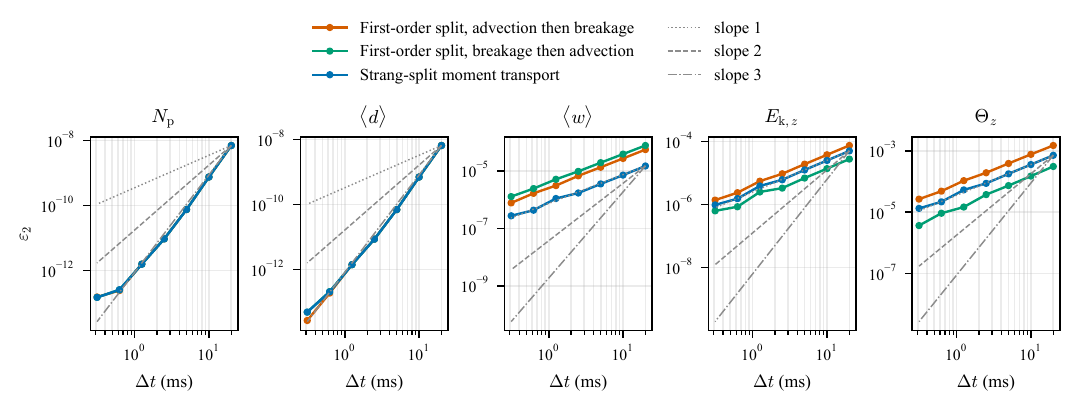}
        \caption{Particle settling and breakage in a flow (Section~\ref{sec:test1}).}
        \label{fig:convergence_test1}
    \end{subfigure}\\[0.5em]
    \begin{subfigure}{\linewidth}
        \includegraphics[width=\linewidth]{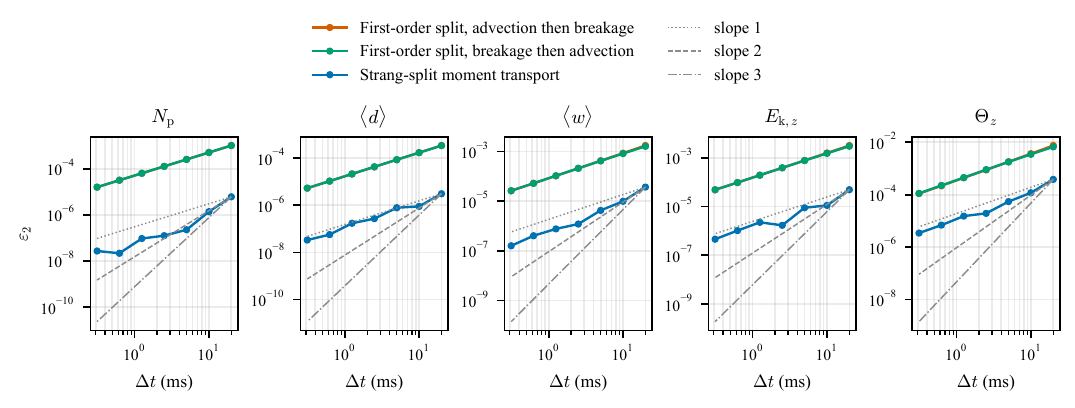}
        \caption{Particle attrition and breakage in a flow (Section~\ref{sec:test3}).}
        \label{fig:convergence_test3}
    \end{subfigure}
    \caption{\reviewertwo{Relative difference in the $L_2$ norm between the CQMOM solutions with successive time steps $\Delta t$ and $\Delta t/2$, for the two first-order splittings, Eqs.~\eqref{eq:operator-splitting-CD} and \eqref{eq:operator-splitting-DC}, and the Strang splitting, Eq.~\eqref{eq:strang}, with $\text{St} = 10$.}}
    \label{fig:convergence}
\end{figure}

\begin{table}[t]
    \centering
    \caption{\reviewertwo{Largest relative deviation (in $\%$) of the velocity statistics (mean velocities, kinetic energies, and granular temperatures) of the CQMOM solution, using the baseline quadrature and Strang splitting, from the mean of the Monte Carlo solution, in the $L_2$ norm, Eq.~\eqref{eq:error_measure_2}, with the maximum norm, Eq.~\eqref{eq:error_measure_inf}, of the same quantity in parentheses, for different time steps $\Delta t$.}}
    \label{tab:time_steps}
    \small
    \reviewertwo{\begin{tabular}{l cc cc cc}
    \hline
    & \multicolumn{2}{c}{Test case 1} & \multicolumn{2}{c}{Test case 2} & \multicolumn{2}{c}{Test case 3} \\
    $\Delta t$ (ms) & $\text{St} = 0.1$ & $\text{St} = 10$ & $\text{St} = 0.1$ & $\text{St} = 10$ & $\text{St} = 0.1$ & $\text{St} = 10$ \\
    \hline
    20 & 15 (17) & 6.8 (6.9) & 12 (13) & 3.9 (6.7) & 11 (13) & 1.8 (1.6) \\
    10 & 32 (17) & 6.8 (6.9) & 2.8 (1.8) & 4.0 (6.8) & 3.8 (0.90) & 1.8 (1.6) \\
    5 & 3.9 (2.7) & 6.9 (7.0) & 3.6 (2.0) & 4.0 (6.8) & 4.3 (0.97) & 1.8 (1.6) \\
    2.5 & $2.3\times 10^{3}$ ($1.6\times 10^{3}$) & 6.9 (7.0) & 2.7 (1.8) & 4.0 (6.8) & 4.4 (0.99) & 1.8 (1.6) \\
    1.25 & $2.5\times 10^{4}$ ($1.3\times 10^{4}$) & 6.9 (7.0) & 3.1 (1.8) & 4.0 (6.8) & 4.4 (1.0) & 1.8 (1.6) \\
    0.625 & 240 (240) & 6.9 (7.0) & 2.8 (1.8) & 4.0 (6.8) & 4.4 (1.0) & 1.8 (1.6) \\
    0.3125 & $1.6\times 10^{5}$ ($1.6\times 10^{5}$) & 6.9 (7.0) & 2.9 (1.8) & 4.0 (6.8) & 5.4 (3.9) & 1.8 (1.6) \\
    0.15625 & $2.8\times 10^{6}$ ($3.3\times 10^{6}$) & 6.9 (7.0) & 2.9 (1.8) & 4.0 (6.8) & 13 (14) & 1.8 (1.6) \\
    \hline
    \end{tabular}}
\end{table}

\reviewertwo{To assess the time integration, all simulations are repeated with the two Lie splittings, Eqs.~\eqref{eq:operator-splitting-CD} and \eqref{eq:operator-splitting-DC}, and with time steps from 20 to 0.156~ms. Figure~\ref{fig:convergence} shows the relative differences between the solutions with successive time steps for the first and third test cases at $\text{St}=10$, and Table~\ref{tab:time_steps} lists the deviations of the velocity statistics from the Monte Carlo solution with Strang splitting for all test cases and time steps.}

\reviewertwo{The breakage rate depends only on the diameter, which the node advection changes only through attrition. In the first two test cases, the particle number and the mean diameter are therefore governed by the breakage source alone and converge with third order, as observed in Figure~\ref{fig:convergence_test1}, the order of accuracy of the SSP-RK scheme used for the integration of the breakage source. In the third test case, attrition couples the particle number and the diameter to the velocity, and they converge like the velocity statistics, see Figure~\ref{fig:convergence_test3}.}

\reviewertwo{For $\text{St}=10$, the velocity statistics converge with first order for all three splittings, with the Strang splitting in a less regular manner. In the first test case, shown in Figure~\ref{fig:convergence_test1}, the differences between successive time steps are of similar magnitude for the three splittings. In the second test case, in which recoil couples the two operators more strongly, the Strang splitting reduces these differences by up to about one order of magnitude compared with the Lie splittings, and in the third test case, shown in Figure~\ref{fig:convergence_test3}, in which attrition couples them further, by one to two orders of magnitude. That the Strang-split solution also converges only with first order indicates a first-order error contribution common to all three splittings. With $\Delta t = 5$~ms, the velocity statistics differ from those with $\Delta t = 2.5$~ms by less than 0.03\,\% with the Strang splitting and by less than 0.4\,\% with the Lie splittings. Their deviations from the Monte Carlo solution change neither with the time step, as seen in Table~\ref{tab:time_steps}, nor by more than one percentage point with the splitting, and are therefore suspected to be caused by the moment closure rather than by the time integration.}

\begin{figure}[t]
    \centering
    \begin{subfigure}{\linewidth}
        \includegraphics[width=\linewidth]{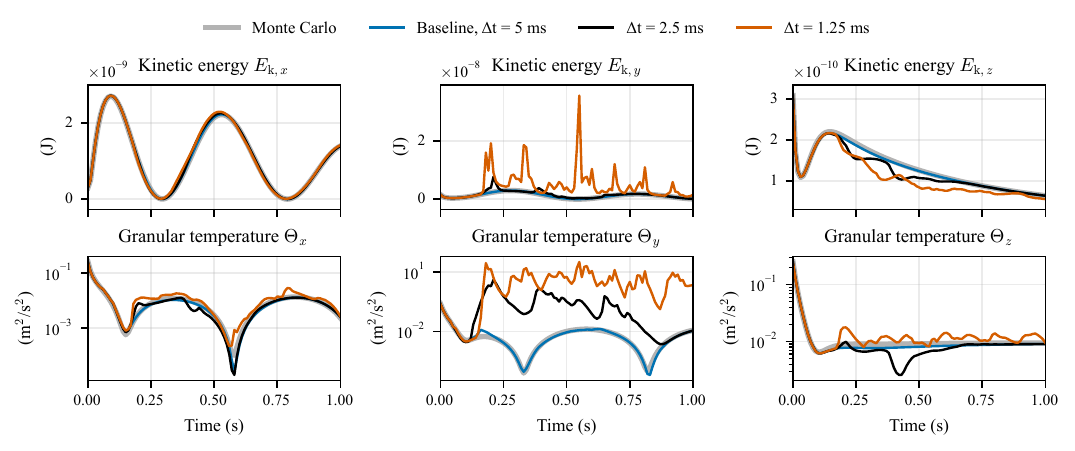}
        \caption{Directional kinetic energies and granular temperatures for different time steps.}
        \label{fig:outliers_physics}
    \end{subfigure}\\[0.5em]
    \begin{subfigure}{\linewidth}
        \includegraphics[width=\linewidth]{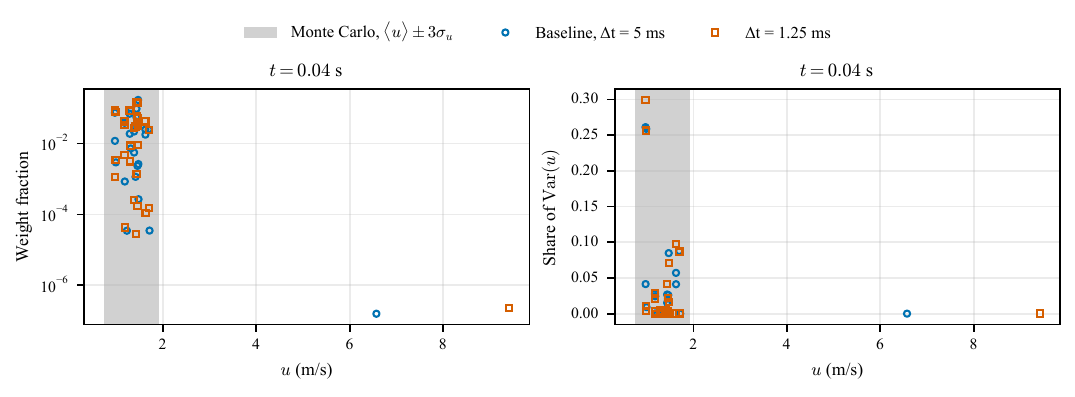}
        \caption{\general{Weight fraction $W_q / \sum_p W_p$ (left) and contribution $W_q (u_q - \langle u \rangle)^2 / \sum_p W_p (u_p - \langle u \rangle)^2$ to the variance of $u$ (right) of each node $\boldsymbol{\xi}_q$ of the quadrature at $t = 0.04$~s.}}
        \label{fig:outliers_nodes}
    \end{subfigure}
    \caption{\reviewertwo{CQMOM results for particle settling and breakage in a flow with $\text{St} = 0.1$ (Section~\ref{sec:test1}) using the baseline quadrature and Strang splitting for different time steps, compared with the Monte Carlo solution. The grey band in (b) is the mean of $u$ of the Monte Carlo solution $\pm$ three standard deviations.}}
    \label{fig:outliers}
\end{figure}

\reviewertwo{For $\text{St}=0.1$, in contrast, the velocity statistics do not converge with decreasing time step for any splitting. As observed in Table~\ref{tab:time_steps}, their deviation from the Monte Carlo solution grows in the first test case from 3.9\,\% with $\Delta t = 5$~ms to more than 100\,\% with $\Delta t = 2.5$~ms and below, remains between 2.7 and 3.6\,\% in the second test case for time steps of 10~ms and below, and grows in the third test case from 4.3\,\% with $\Delta t = 5$~ms to 13\,\% with $\Delta t = 0.156$~ms. Figure~\ref{fig:outliers} shows the cause for the first test case. With $\Delta t = 2.5$~ms and $\Delta t = 1.25$~ms, $\Theta_y$ exceeds the Monte Carlo value by up to several orders of magnitude, as observed in Figure~\ref{fig:outliers_physics}, driven by individual nodes of tiny weight at physically implausible velocities. At $t = 0.04$~s, for example, a branch of $u$ with a standard deviation of about 1~cm/s contains a node at $u \approx 6.6$~m/s with a time step of $\Delta t = 5$~ms and at $u \approx 9.4$~m/s with $\Delta t = 1.25$~ms, each carrying of the order of $10^{-7}$ of the population as seen in Figure~\ref{fig:outliers_nodes}. With $\Delta t = 5$~ms, the branch of $v$ conditioned on this node contains a node at about $60$~m/s. A two-node quadrature reproduces the mean, the variance, and the third moment of its branch, so that in a branch with a very small standard deviation even a small excess of the third moment can only be reproduced by a distant node of small weight, as quantified in Section~\ref{sec:discussion_scope}. Since on the real line any third moment is admissible once the variance is positive, and the same holds for the highest moment of a quadrature with more nodes, the corresponding moment set remains realizable. Relative to its branch, the outlier node carries about $6 \times 10^{-6}$ of the weight and thus passes the weight-ratio threshold of $10^{-6}$ of the adaptive Wheeler algorithm. Such outlier nodes arise at every time step, but they grow with the number of quadrature reconstructions per unit time, so that smaller time steps amplify them. The temporary overshoot of $\Theta_y$ at $t \approx 0.16$~s in the first test case with $\Delta t = 5$~ms, see Figure~\ref{fig:test1_st01_physics}, is an early stage of the same mechanism: nodes more than five standard deviations from the mean of their branch, carrying less than $10^{-3}$ of the population, account for more than half of the variance of $v$, and without them the variance agrees with the Monte Carlo solution to within about 1\,\%.}

\reviewertwo{Larger time steps, in turn, increase the splitting error: with $\Delta t = 20$~ms, the deviations for $\text{St}=0.1$ reach 11--15\,\%. The time step of $\Delta t = 5$~ms used in Section~\ref{sec:results} is therefore a compromise rather than a converged choice. For the same reason, the larger deviations of the Lie splittings for $\text{St}=0.1$ and $\Delta t = 5$~ms, in the first test case 31\,\% with advection first and 13\,\% with breakage first, compared with 3.9\,\% with Strang splitting, cannot be attributed to the splitting error alone.}

\subsection{Quadrature resolution and conditioning order}
\label{sec:discussion_resolution}

\begin{table}[t]
    \centering
    \caption{\reviewertwo{Largest relative deviation (in $\%$) of the velocity statistics (mean velocities, kinetic energies, and granular temperatures) of the CQMOM solution, using Strang splitting and $\Delta t = 5$~ms, from the mean of the Monte Carlo solution, in the $L_2$ norm with the maximum norm of the same quantity in parentheses, as in Table~\ref{tab:time_steps}, for different quadratures $(N_d, N_\chi, N_u, N_v, N_w)$.}}
    \label{tab:resolution}
    \footnotesize
    \setlength{\tabcolsep}{4pt}
    \reviewertwo{\begin{tabular}{l cc cc cc}
    \hline
    & \multicolumn{2}{c}{Test case 1} & \multicolumn{2}{c}{Test case 2} & \multicolumn{2}{c}{Test case 3} \\
    Quadrature & $\text{St} = 0.1$ & $\text{St} = 10$ & $\text{St} = 0.1$ & $\text{St} = 10$ & $\text{St} = 0.1$ & $\text{St} = 10$ \\
    \hline
    $(2,2,2,2,2)$ & 3.9 (2.7) & 6.9 (7.0) & 3.6 (2.0) & 4.0 (6.8) & 4.3 (0.97) & 1.8 (1.6) \\
    $(3,2,2,2,2)$ & 200 (110) & $1.3\times 10^{6}$ ($1.8\times 10^{6}$) & 3.5 (1.8) & $2.9\times 10^{4}$ ($4.2\times 10^{4}$) & 3.2 (0.90) & 350 (510) \\
    $(2,3,2,2,2)$ & 4.9 (2.2) & 6.9 (7.0) & 2.7 (1.7) & 4.0 (6.8) & 3.1 (1.4) & 1.8 (1.6) \\
    $(3,3,2,2,2)$ & 77 (37) & $4.6\times 10^{5}$ ($1.8\times 10^{6}$) & 2.8 (1.7) & 5.8 (8.5) & 3.1 (0.80) & $1.5\times 10^{4}$ ($2.6\times 10^{4}$) \\
    $(2,2,3,3,3)$ & $3.4\times 10^{3}$ ($3.2\times 10^{3}$) & 6.9 (7.0) & 2.6 (0.61) & $1.5\times 10^{5}$ ($3.7\times 10^{5}$) & 2.5 (0.67) & $1.4\times 10^{3}$ ($2.2\times 10^{3}$) \\
    $(3,3,3,3,3)$ & 8.8 (4.3) & $9.5\times 10^{4}$ ($1.2\times 10^{5}$) & 3.6 (2.3) & $2.6\times 10^{4}$ ($5.3\times 10^{4}$) & 2.5 (0.71) & $2.5\times 10^{4}$ ($5.2\times 10^{4}$) \\
    \hline
    \end{tabular}}
\end{table}

\reviewertwo{To assess the influence of the quadrature resolution on the results, all simulations are performed with six quadratures, from $2^5$ to $3^5$ nodes, listed in Table~\ref{tab:resolution}. For $\text{St}=0.1$, the velocity statistics of the second and third test cases deviate from the Monte Carlo solution by 2.5--4.3\,\% for all quadratures, whereas in the first test case half of the tested quadratures produce deviations larger than 70\,\%. For $\text{St}=10$, increasing the quadrature resolution of the aspect-ratio coordinate leaves the velocity statistics largely unchanged, whereas increasing the quadrature resolution of the diameter or the velocity components produces large deviations in most cases, through the same outlier nodes as at small time steps. Refining the quadrature therefore does not reduce the deviations systematically, which we regard as a limitation of the conditional velocity hierarchy rather than of the quadrature resolution. The deviation of the variance of the aspect ratio is limited by $N_d$ rather than by $N_\chi$, because breakage correlates the aspect ratio with the diameter: $N_\chi = 3$ leaves it unchanged, whereas $N_d = 3$ reduces it from 5.1--26\,\% to 2.4--5.0\,\% in the first two test cases and from 5.1 to 1.8\,\% in the third test case for $\text{St}=10$, but increases it from 0.2 to 5.0\,\% in the third test case for $\text{St}=0.1$. The baseline nevertheless uses $N_d = 2$, since the quadrature $(3,2,2,2,2)$ produces large deviations of the velocity statistics in most cases.}

\begin{figure}[t]
    \centering
    \includegraphics[width=\linewidth]{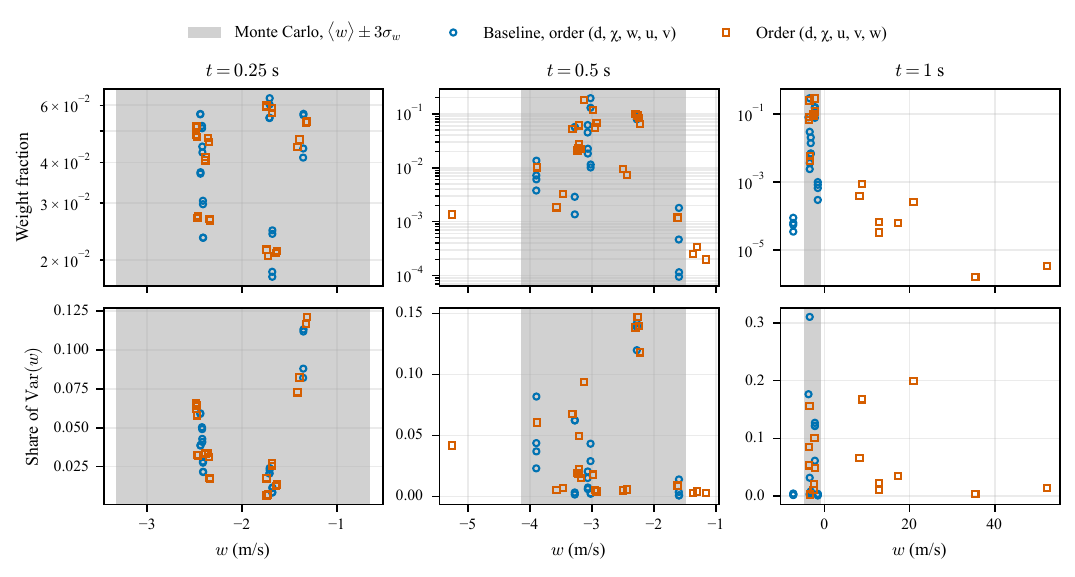}
    \caption{\reviewerone{CQMOM results for particle attrition and breakage in a flow with $\text{St} = 10$ (Section~\ref{sec:test3}) using the baseline quadrature and Strang splitting in the baseline conditioning order $(d, \chi, w, u, v)$ and in the natural order $(d, \chi, u, v, w)$: weight fraction (top) and contribution to the variance of $w$ (bottom), defined as in Figure~\ref{fig:outliers}, of each node $\boldsymbol{\xi}_q$ of the quadrature. The grey band is the mean of $w$ of the Monte Carlo solution $\pm$ three standard deviations.}}
    \label{fig:order}
\end{figure}

\reviewerone{The conditioning order matters as well. In the natural order $(d, \chi, u, v, w)$, the vertical velocity $w$ is conditioned last, on the horizontal velocities $u$ and $v$. As observed in Figure~\ref{fig:order}, in the third test case for $\text{St}=10$, the quadrature of $w$ then develops nodes at $w \approx 10$--$50$~m/s, far outside the Monte Carlo population, which together carry about half of the variance of $w$ at $t = 1$~s. As a result, the $L_2$ norm of $\Theta_z$ deviates from the Monte Carlo solution by 31\,\%, with its maximum norm deviating by 83\,\%. Conditioning $w$ before $u$ and $v$, as in the quadrature used for the results in Section \ref{sec:results}, keeps the nodes of $w$ close to the Monte Carlo population and reduces this deviation to 0.9\,\% (1.1\,\%). In the other test cases, the largest deviation of the velocity statistics changes by at most one percentage point between the two orders, although individual quantities change by up to two percentage points.}

\subsection{Realizability and conditioning of the quadrature reconstruction}
\label{sec:discussion_realizability}

\begin{figure}[t]
    \centering
    \includegraphics[width=\linewidth]{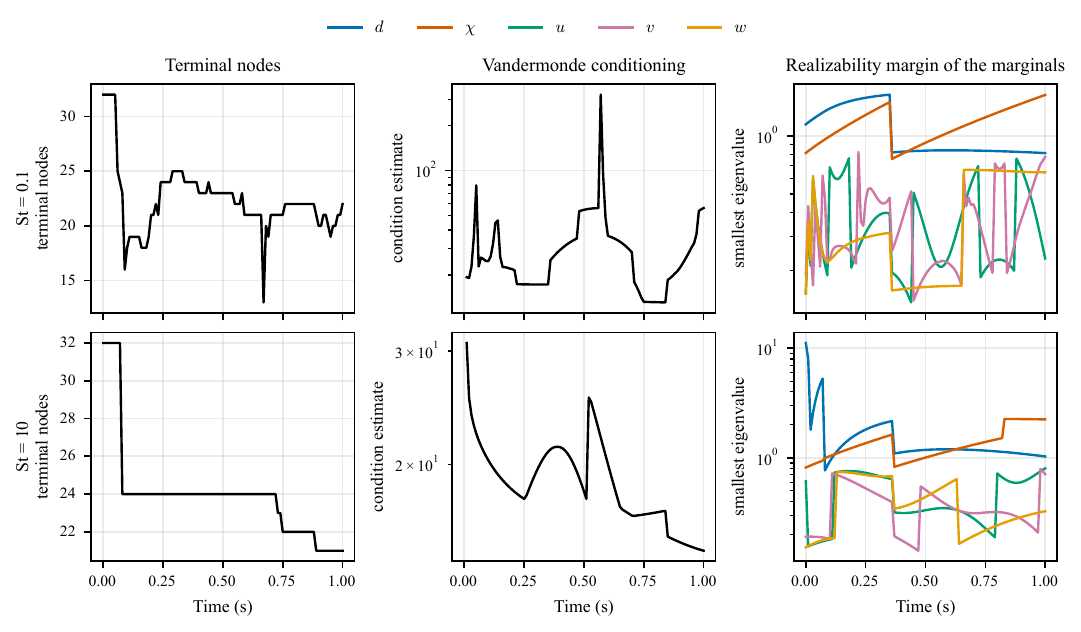}
    \caption{\reviewertwo{CQMOM results for particle attrition and breakage in a flow (Section~\ref{sec:test3}) using the baseline quadrature and Strang splitting: number of quadrature nodes, largest condition estimate of the conditional Vandermonde systems in the reconstructions since the previous output time, and smallest eigenvalue of the standardized Hankel matrices of the marginal moments of each coordinate, including those that account for its support, which vanishes at the boundary of the moment space.}}
    \label{fig:reconstruction}
\end{figure}

\reviewertwo{No realizability test failed and no Vandermonde system was refused in any of the simulations reported in this work. Figure~\ref{fig:reconstruction} shows, for the third test case, the number of quadrature nodes, the condition estimates of the conditional Vandermonde systems, and the smallest eigenvalue of the standardized Hankel matrices of each marginal, including those of the shifted moment sequences that account for the bounded support of $d$ and $\chi$, which vanishes at the boundary of the moment space. In all baseline simulations, this eigenvalue remains between 0.13 and 11, far from zero, and the condition estimates remain between 11 and $1.4 \times 10^3$. In the baseline simulations of all test cases, the adaptive quadrature reconstruction reduces the number of quadrature nodes from 32 to as few as 8 at $\text{St}=0.1$ (first test case) and 21 at $\text{St}=10$ (third test case), and every reduction in the reconstruction after the integration of the breakage source projects the moments onto those reproduced by the reduced quadrature. The brief dips of $\Theta_y$ in the second test case at $\text{St}=0.1$, see Figure~\ref{fig:test2_st01_physics}, coincide with bursts of such reductions, in which a velocity branch reduced to a single node loses its conditional variance, which then recovers over the following time steps. The deviations of Section~\ref{sec:results} and the outlier nodes are therefore caused neither by a loss of realizability nor by ill-conditioned Vandermonde systems, but by the closure itself: a realizable conditional velocity quadrature may contain nodes of tiny weight at physically implausible velocities.}

\subsection{Computational cost}
\label{sec:discussion_cost}

\reviewertwo{On a single core (Apple M2 Pro, Apple clang~21 with \texttt{-O3}), the CQMOM simulation of 1~s of physical time with the baseline quadrature of two nodes per coordinate and a time step of $\Delta t = 5$~ms takes 0.07--0.09~s, the shortest of three runs, compared with 5.0--6.1~s for one Monte Carlo realization of initially $10^4$ particles. With $3^5$ nodes, the CQMOM simulation takes 0.55--1.03~s. All three splittings perform four CQMOM inversions per time step, so that the Strang splitting, which only divides the node advection into two half steps, costs 3--19\,\% more than the Lie splittings. For a Strang step, the CQMOM quadrature reconstruction together with the realizability tests and the Runge-Kutta updates of the moments in the integration of the breakage source takes 50--60\,\%, the evaluation of the breakage rates 19--25\,\%, and the node advection 18--31\,\%. The share of the node advection is largest for $\text{St}=0.1$, where the adaptive integration of the node trajectories requires 12--20 substeps per time step, compared with 6 substeps for $\text{St}=10$.}

\subsection{Scope of the particle model and extensions}
\label{sec:discussion_scope}

\reviewerthree{The particle Reynolds numbers of the Monte Carlo populations have medians of 3--4 for $\text{St}=0.1$, with 95th percentiles of up to 30, and of 200--250 for $\text{St}=10$, with 95th percentiles of up to about 380. The populations with $\text{St}=10$ are therefore in the range in which freely moving disks and oblate spheroids may flutter or tumble \citep{Ern2012, Zhou2017, Tinklenberg2023}, which the QSO assumption of the particle model excludes. More generally, the QSO assumption requires that the pitching torque aligns a particle faster than its orientation is perturbed. Non-spherical particles in turbulent flows tend to maximize their drag, but their inertia and the local velocity fluctuations and gradients prevent this from occurring instantaneously \citep{VanWachem2015}, and collisions with other particles or walls perturb their orientation further. As discussed in Section~\ref{sec:model}, the test cases scrutinize the numerical method for a given particle model, which the Monte Carlo solution shares, but they do not validate the particle model. Relaxing the QSO assumption would require resolving the orientation and angular velocity of the particles, which extends the phase space to seven dimensions for in-plane pitching (Section~\ref{sec:eom}), as well as closures for the pitching torque and the rotational damping over the Reynolds numbers of interest.}
\reviewerone{In addition, the particle populations considered here are dilute and collisionless. Including collisions in a physically meaningful manner would not be straightforward, since the collision kernels of non-spherical particles depend on the relative orientation, the contact point, and the angular velocities of the colliding particles \citep{Siewert2014, Kheradmand2022, Vescovi2024}.}

\reviewerone{The outlier nodes of the conditional velocity quadratures, discussed in Section~\ref{sec:discussion_splitting}, identify the representation of the velocity distribution as the main weakness of the present formulation. On the bounded supports of $d$ and $\chi$, a node outside the support reduces the node count \citep{Denner2026a}, whereas no such bound exists for the velocities. For a two-node quadrature of a distribution with standard deviation $\sigma$ and normalized skewness $s$, the nodes lie at $\sigma\,(s \pm \sqrt{s^2+4})/2$ from the mean with weights $(1 \mp s/\sqrt{s^2+4})/2$, as follows from \citet[Section 2]{Fox2018} with the normalized flatness $\eta = 1 + s^2$. Hence, for $s \gg 1$, one node carries a weight of approximately $1/s^2$ and lies at a distance of approximately $\sigma s = C_3/C_2$ from the mean, where $C_2$ and $C_3$ are the second and third central moments of the branch. This distance diverges as the branch becomes monokinetic at a fixed third moment. The outlier node of Section~\ref{sec:discussion_splitting}, which carries about $6 \times 10^{-6}$ of the weight of its branch, corresponds to $s \approx 400$. The weight-ratio threshold of the adaptive Wheeler algorithm limits $s$ to about $10^3$, but does not prevent such nodes. To remedy this weakness, the velocity distribution within each branch of the CQMOM quadrature for $d$ and $\chi$ may be represented with the conditional hyperbolic quadrature method of moments (CHyQMOM) \citep{Fox2018, Patel2019a}. In CHyQMOM, the conditional mean and variance depend linearly on the conditioning velocity, whereas the normalized skewness and flatness are shared by all conditional branches and are determined from the joint moments. The distance of the nodes from the mean of a branch, in units of its standard deviation, is thus set by the joint distribution rather than by the moments of the individual branch, whose normalized skewness can become extreme if its standard deviation is very small. Moreover, the largest abscissa can be bounded by limiting the skewness and flatness, at the cost of no longer reproducing the third- and fourth-order moments \citep{Patel2019a}. CHyQMOM also ensures the hyperbolicity of the moment fluxes required for coupling with transport in physical space.}

\section{Conclusions}
\label{sec:conclusions}

This work presented a five-dimensional conditional quadrature method of moments (CQMOM) for the simulation of non-spherical particle populations. The proposed methodology combines a conditional quadrature representation of the particle size, aspect ratio, and velocity of oblate spheroids with a quasi-steady orientation assumption, reducing the dimensionality of the underlying phase space while retaining the influence of particle shape on the aerodynamic forces. To facilitate the prediction of coupled geometric and kinematic particle dynamics, Strang splitting was used to separate continuous and discontinuous processes in phase space.

The proposed five-dimensional CQMOM framework was \general{compared} against Monte Carlo simulations \general{of the same particle model} for three representative test cases of increasing complexity, involving aerodynamic transport, binary breakage, stochastic recoil, and surface attrition\general{, for particle populations with Stokes numbers of 0.1 and 10}. \general{With two nodes per coordinate, the CQMOM solution reproduces the particle number, the mean diameter, and the mean aspect ratio within the statistical uncertainty of the Monte Carlo solution, the mean velocities and kinetic energies to within 1.8\,\%, and the granular temperatures to within 6.9\,\%.} The results show that the employed conditional quadrature representation is capable of preserving the geometric-kinematic correlations, despite the high dimensionality of the underlying phase space. \general{For $\text{St} = 10$, Strang splitting reduces the time-step error by up to two orders of magnitude compared with first-order Lie splitting. The time-step and resolution studies also reveal the main limitation of the present formulation: in nearly degenerate conditional velocity distributions, the quadrature may contain nodes of tiny weight at physically implausible velocities, thus identifying avenues for further research.}

The proposed framework establishes a foundation for the Euler-Euler simulation of polydisperse \general{oblate spheroidal} particle populations and extends the applicability of CQMOM to an important class of industrially relevant disperse multiphase flows. More generally, the present work demonstrates that conditional quadrature methods can be extended to high-dimensional phase spaces involving coupled geometric and kinematic particle properties.

\section*{Acknowledgements}
\noindent We gratefully acknowledge the financial support of the Natural Sciences and Engineering Research Council of Canada (NSERC), funding reference number ALLRP-592877-24, PRIMA Québec, and Nouveau Monde Graphite.

\section*{Data availability}
\noindent The presented results as well as the computer code, configuration files, and scripts required to reproduce all results, figures, and tables of this work are available in Zenodo at \url{https://doi.org/10.5281/zenodo.24253462}.

\section*{Declaration of generative AI and AI-assisted technologies}
\noindent During the preparation of this work the authors used ChatGPT (OpenAI) and Claude (Anthropic) in order to improve the readability of the manuscript. After using these tools, the authors reviewed and edited the content as needed and take full responsibility for the content of the published article.

\end{document}